\documentclass[runningheads]{llncs}

\usepackage[T1]{fontenc}
\usepackage{graphicx}
\usepackage{amsmath,amssymb}
\usepackage{hyperref}
\usepackage{url}
\usepackage{algorithm}
\usepackage{todonotes}
\usepackage{xcolor}
\usepackage{tikz}
\usetikzlibrary{positioning,arrows.meta,calc,shapes.symbols,decorations.markings,decorations.pathreplacing,fit}
\usepackage{comment}
\usepackage{algpseudocode}
\usepackage{booktabs}
\usepackage{array}
\usepackage{placeins}
\usepackage{hyperref}

\begin{document}

\title{The Cost of Stability: Deanonymizing Onion Services Through Long-Lived Introduction Circuits}
\titlerunning{Deanonymizing Onion Services Through Long-Lived Introduction Circuits}

\author{%
  \begin{tabular}{c@{\hspace{4em}}c}
    Nicolas Constantinides$^{*}$ & Mahdi Rahimi \\[2pt]
    \normalsize Unaffiliated     & \normalsize COSIC, KU Leuven \\
                                 & \normalsize Leuven, Belgium
  \end{tabular}}
\authorrunning{N. Constantinides and M. Rahimi}
\institute{}
\maketitle
\begin{comment}
\begin{abstract}
Tor hides communication paths by routing traffic through circuits of relays.
Identifying the next relay from network traffic is difficult because each relay
carries traffic for many circuits at once. Ordinary circuits also last only
about $10$~min, limiting observations of a fixed path.

Onion services, however, use \emph{introduction circuits} that remain fixed for
$18$--$24$~h. We exploit this long lifetime with an adaptive intersection attack
that observes only one relay at a time. Starting from the publicly advertised
relay at the client-facing end of the circuit, the adversary repeatedly triggers
introductions and records the destinations contacted by that relay. Intersecting
these observations isolates its successor, which is then observed in the same
way. Repeating this process reconstructs the circuit hop by hop and ultimately
reveals the onion service's network location.

We evaluate the attack on the live Tor network against a self-operated onion
service carrying genuine third-party traffic. Across nine end-to-end experiments,
the attack reconstructs the complete path in every run, with an estimated median
time of $3.2$~h. We propose periodically rebuilding the internal introduction path to
limit evidence accumulation while keeping the publicly advertised relay
unchanged.
\keywords{Tor \and onion services \and intersection attack \and traffic analysis}
\end{abstract}
\end{comment}

\begin{abstract}

Tor is a widely used anonymity network that provides network privacy by
routing client communications through a sequence of relays to their
destinations. An adversary observing a single Tor relay cannot readily
link clients to their destinations, as doing so requires identifying the
other relays along a client's circuit. Identifying these relays from traffic
observed at the monitored relay alone is challenging: each relay carries
traffic for many circuits simultaneously, and client circuits typically
last only about $10$~min, limiting observations of a given circuit.

In contrast, Tor onion services use \emph{introduction circuits} that
remain fixed for $18$--$24$~h. We develop an intersection attack that
exploits this design, allowing an adaptive adversary observing one Tor
relay at a time to identify an onion service's network location.
We evaluate the attack on the live Tor network against a self-operated
onion service carrying genuine third-party traffic. Across nine
end-to-end experiments, the attack reconstructs the complete Tor
circuit in every run, with an estimated median time of $2.2$~h.
To mitigate this attack, we propose a mechanism that periodically
rebuilds internal introduction circuits to limit evidence accumulation
without changing publicly advertised information.


\end{abstract}

\section{Introduction}

Tor~\cite{dingledine2004tor} is the most widely used anonymity network,
enabling millions of users to conceal their sensitive Internet activities
from network routers, Internet service providers (ISPs), and potentially
nation-state adversaries. Tor is a low-latency anonymity network that
provides network privacy by routing traffic through a \emph{circuit} of
relays using layered encryption before forwarding it to its destination.
Each relay along the communication path removes one layer of encryption
and learns only its immediate predecessor and successor, preventing any
single relay from identifying both the client and its destination, thereby
preserving communication anonymity.

Tor's privacy protections also extend to service operators through
\emph{onion services}, historically referred to as \emph{hidden services}
\cite{huete2021tor}. Onion services allow clients and servers to communicate
without directly revealing their respective network locations to one another,
enabling applications that require privacy for both parties
\cite{avarikioti2018structure}. Since their introduction, onion services
have been the subject of extensive research into attacks that seek to
identify their operators
\cite{biryukov2013trawling,jansen2018inside,kwon2015circuit,ling2013protocol,platzer2020critical}.

Tor's objective is to limit \emph{traffic analysis}, in which an adversary
uses traffic patterns observed at a monitored relay to determine which network
point to attack next~\cite{dingledine2004tor}. For an adversary limited to monitoring one relay at a time, however, such
attacks face a central challenge. Starting from one end of the targeted onion
service's circuit, the adversary must identify the next relay on that circuit
among the many circuits the monitored relay carries concurrently, then move its
observation to that relay and repeat, one hop at a time, until the circuit is
reconstructed. Successfully identifying the relevant circuit
can enable the adversary to determine the onion service's network location.
This task is particularly challenging for ordinary Tor circuits, which
typically last at most about $10$~min, limiting the evidence that can be
collected about a given circuit \cite{erdin2015find}. Once a circuit
terminates, its state is discarded, so compromising a relay afterward
does not reveal its previous circuit state \cite{basyoni2020traffic}.
An adversary with only a partial view of the network must therefore
identify the relevant relays before the circuit terminates.

However, such attacks can be particularly effective against onion services,
as each \emph{introduction circuit} of an onion service remains active for
$18$--$24$~h \cite{platzer2020critical}. An onion service maintains such a
long-lived circuit to each of its introduction points, through which clients
initiate contact while the service's network location remains concealed.
This substantially longer lifetime gives an adaptive adversary that observes
only one Tor relay at a time an extended window in which to accumulate
evidence about the circuit's path, raising the possibility that observations
too weak to identify a short-lived circuit become decisive when accumulated
over the lifetime of an introduction circuit.

\noindent
\paragraph{\textbf{Threat model and objective.}}
We consider an adversary that monitors one relay at a time, chosen adaptively:
it observes the traffic passing through the monitored relay and uses those
observations to identify the next relay on the circuit, then moves on. Repeating
this, it aims to identify the relays leading to the onion service.
Motivated by the unusually long lifetime of introduction circuits, our goal
is to investigate whether such an adversary can accumulate sufficient
evidence over time to ultimately identify the onion service's network
location, thereby compromising the privacy guarantees provided by Tor. We
then investigate how to mitigate this attack by reducing the lifetime of
introduction circuits.

\noindent
\paragraph{\textbf{Contributions.}}
This paper makes the following contributions:

\noindent
\textbf{(1).}
We identify the long lifetime of onion-service introduction circuits as a
deanonymization risk under \emph{partial, sequential} network observation
and formalize an adaptive intersection attack that reconstructs a fixed
introduction circuit hop by hop from repeated introduction-handshake
observations. This demonstrates how traffic analysis can be performed within
the partial-observation setting that Tor is designed to resist
(Section~\ref{sec:attack}).

\noindent
\textbf{(2).}
We evaluate the attack end to end against a self-operated onion service on
the live Tor network under genuine third-party background traffic, following
the safeguards of the Tor Research Safety Board. The attack reconstructs the
complete target path in all nine experiments (Section~\ref{sec:eval}).

\noindent
\textbf{(3).}
We propose a mitigation that reduces the lifetime of introduction circuits
by periodically rebuilding the internal service-side path while preserving
the Introduction Point. This limits the attack without changing Tor's
publicly visible information (Section~\ref{sec:mitigation}).

\section{Background}

\label{sec:background}

This section explains how Tor builds circuits and how a client connects to an onion
service. Both are needed to follow the attack in Section~\ref{sec:attack}.

\subsection{Tor Circuits and Path Selection}

\label{sec:path-selection}

Tor provides network privacy for users by relaying traffic through a \emph{circuit} of
volunteer-run \emph{relays}, carried in fixed-size messages called
\emph{cells}. A circuit is built by adding relays one by one, and each relay
knows only the relay before and after it on the circuit. Relays are chosen from
the \emph{consensus}, a public list of all relays. The chance that a relay is
chosen is proportional to its \emph{consensus weight} (CW), a value based on
its bandwidth, so higher-weight relays carry more traffic. In particular, the
entry relay of the circuit, the \emph{guard} relay, in Tor is used for a long
time for multiple circuits of a user.\footnote{\href{https://spec.torproject.org/guard-spec/index.html}{Tor
Specifications, ``Tor Guard Specification.''}} In onion services, a similar
principle, known as \emph{Vanguards-Lite}, is enabled by
default.\footnote{\href{https://spec.torproject.org/vanguards-spec/index.html}{Tor
Specifications, ``Tor Vanguards Specification.''}} The second relay for
onion services is chosen from a fixed set of four \emph{layer-2 vanguards},
each kept for a random period of $1$--$12$~days. The optional \emph{Full
Vanguards} mode also fixes the third relay. As a result, the relays closest to
an onion service change slowly, even when the service builds many circuits.

\subsection{Onion Service Protocol}

\label{sec:onion-service}

% Per-circuit parallel-lane variant: each circuit's request/ack/forward/app lines
% run as their own constant-spacing parallel lanes (per-vertex miter offset).
% Single shared entry guard per side: the guard sits on its own, further from the
% vanguard column, drawn larger so it reads as the hop all three circuits share.
\definecolor{stepone}{HTML}{332288}
\definecolor{steptwo}{HTML}{EE3377}
\definecolor{stepthree}{HTML}{CC3311}
\definecolor{stepfour}{HTML}{0077BB}
\definecolor{stepfive}{HTML}{33BBEE}
\definecolor{stepsix}{HTML}{EE7733}
\definecolor{stepseven}{HTML}{117733}
\definecolor{stepeight}{HTML}{009988}
\definecolor{stepnine}{HTML}{444444}
\begin{figure*}[!t]
\centering
\resizebox{\linewidth}{!}{%
\begin{tikzpicture}[
  >={Latex[length=3mm,width=2mm]},
  cas/.style={preaction={draw,line width=4.4pt,white,rounded corners=3pt}},
  req/.style={->,line width=2.3pt,rounded corners=3pt,cas},
  fwd/.style={->,line width=2.3pt,dash pattern=on 3.6pt off 2.4pt,rounded corners=3pt,cas},
  ackl/.style={->,line width=1.8pt,densely dotted,preaction={draw,line width=3.5pt,white}},
  appl/.style={{Latex[length=3.2mm,width=2.6mm]Latex[length=3.2mm,width=2.6mm,reversed,sep=2.4mm]}-{Latex[length=3.2mm,width=2.6mm,reversed,sep=2.4mm]Latex[length=3.2mm,width=2.6mm]},line width=2.5pt,rounded corners=3pt,cas},
  rb/.style n args={2}{circle,fill=#1,text=#2,font=\normalsize\bfseries,inner sep=0.3pt,minimum size=6.0mm,preaction={draw=white,line width=3pt}},
  rbs/.style n args={2}{circle,fill=#1,text=#2,font=\scriptsize\bfseries,inner sep=0.2pt,minimum size=4.2mm,preaction={draw=white,line width=1.6pt}},
  abd/.style={circle,draw=#1,dash pattern=on 0.4pt off 1.4pt,line width=0.9pt,fill=white,text=#1!80!black,font=\normalsize\bfseries,inner sep=0.3pt,minimum size=6.2mm,preaction={draw=white,line width=2.5pt}},
  stn/.style={fill=white,draw=gray!45,line width=0.5pt},
  rlab/.style={font=\large\bfseries,inner sep=1.5pt,fill=white,rounded corners=1pt},
  elab/.style={font=\large\bfseries,align=center,inner sep=1.5pt}
]
% ---------------- coordinates ----------------
\coordinate (hs) at (-1.9,0);   \coordinate (cl) at (22.6,0);
% one shared entry guard per side, vertically centred on the vanguard column
\coordinate (egS) at (2.2,0.0);   \coordinate (egC) at (18.8,0.0);
\coordinate (vgSh) at (5.8,2.8);  \coordinate (vgSi) at (5.8,0.0);  \coordinate (vgSr) at (5.8,-2.8);
\coordinate (vgCh) at (15.2,2.8); \coordinate (vgCi) at (15.2,0.0); \coordinate (vgCr) at (15.2,-2.8);
\coordinate (vgS4) at (5.8,-5.3);  \coordinate (vgC4) at (15.2,-5.3); % 4th (unused) L2 vanguards
\coordinate (mSd) at (6.8,6.3);  \coordinate (mSi) at (8.0,3.2);  \coordinate (mSr) at (8.0,-3.2);
\coordinate (mCd) at (14.2,6.3); \coordinate (mCi) at (13.0,3.2); \coordinate (mCr) at (13.0,-3.2);
\coordinate (ip) at (10.5,5.9);  \coordinate (rp) at (10.5,-5.9); \coordinate (hd) at (10.5,9.2);
\node[cloud, cloud puffs=22, cloud puff arc=104, aspect=1.9, draw=gray!30,
      line width=0.4pt, fill=gray!7, minimum width=21.6cm, minimum height=15.5cm]
      at (10.5,0.1) {};
% generated parallel-lane draws (per-vertex miter offset, delta=0.34)
\draw[req,steptwo] (-0.550,0.820) -- (1.900,0.820) -- (2.545,0.161) -- (5.882,2.742) -- (6.882,6.243) -- (9.767,8.520);
\draw[ackl,steptwo] (10.269,9.449) -- (6.312,6.606) -- (5.335,3.258) -- (2.388,0.394) -- (1.900,1.000) -- (-0.550,1.000);
\draw[req,stepone] (-0.550,0.000) -- (1.900,0.000) -- (2.500,0.045) -- (5.800,0.000) -- (8.000,3.200) -- (9.685,5.019);
\draw[ackl,stepone] (10.251,6.131) -- (7.734,3.413) -- (5.621,0.340) -- (2.500,0.000) -- (1.900,0.180) -- (-0.550,0.180);
\draw[fwd,stepsix] (10.749,5.669) -- (8.266,2.987) -- (5.979,-0.340) -- (2.500,-0.045) -- (1.900,-0.180) -- (-0.550,-0.180);
\draw[req,stepseven] (-0.550,-0.820) -- (1.900,-0.820) -- (2.500,-0.205) -- (5.800,-2.800) -- (8.000,-3.200) -- (9.685,-5.019);
\draw[req,stepthree] (21.250,1.000) -- (19.100,1.000) -- (18.500,0.250) -- (15.200,2.800) -- (14.200,6.300) -- (11.233,8.520);
\draw[ackl,stepthree] (10.269,8.951) -- (13.888,6.126) -- (14.921,2.460) -- (18.500,0.205) -- (19.100,0.820) -- (21.250,0.820);
\draw[req,stepfive] (21.250,0.180) -- (19.100,0.180) -- (18.500,0.180) -- (15.200,0.300) -- (13.000,3.200) -- (11.315,5.019);
\draw[ackl,stepfive] (10.251,5.669) -- (12.734,2.987) -- (15.021,-0.340) -- (18.500,0.000) -- (19.100,0.000) -- (21.250,0.000);
\draw[req,stepfour] (21.250,-0.820) -- (19.100,-0.820) -- (18.500,-0.160) -- (15.200,-2.800) -- (13.000,-3.200) -- (11.315,-5.019);
\draw[fwd,stepeight] (10.749,-6.131) -- (13.172,-3.514) -- (15.343,-3.281) -- (18.612,-0.391) -- (19.100,-1.000) -- (21.250,-1.000);
\draw[ackl,stepfour] (10.251,-5.669) -- (12.828,-2.886) -- (15.057,-2.319) -- (18.388,-0.064) -- (19.100,-0.640) -- (21.250,-0.640);
\draw[appl,stepnine] (-0.550,-1.000) -- (1.900,-1.000) -- (2.500,-0.250) -- (5.739,-3.480) -- (7.655,-3.828) -- (10.500,-6.901) -- (13.345,-3.828) -- (15.398,-3.652) -- (18.637,-0.602) -- (19.100,-1.180) -- (21.250,-1.180);
% badges
\node[rb={steptwo}{white}] at (4.150,1.402) {2a};
\node[abd=steptwo] at (8.023,8.057) {2b};
\node[rbs={stepone}{white}] at (6.900,1.600) {1a};
\node[abd=stepone] at (8.992,4.772) {1b};
\node[rb={stepsix}{white}] at (9.508,4.328) {6};
\node[rb={stepseven}{white}] at (4.150,-1.503) {7};
\node[rb={stepthree}{white}] at (16.850,1.525) {3a};
\node[abd=stepthree] at (12.328,7.038) {3b};
\node[rb={stepfive}{black}] at (16.700,0.245) {5a};
\node[abd=stepfive] at (11.492,4.328) {5b};
\node[rb={stepfour}{white}] at (16.850,-1.480) {4a};
\node[rb={stepeight}{white}] at (11.961,-4.823) {8};
\node[abd=stepfour] at (11.539,-4.277) {4b};
\node[rb={stepnine}{white}] at (9.000,-5.150) {9};

% ---------------- vanguard / middle-relay nodes --------------------
\foreach \c/\l in {vgSh/VG-S-1,vgSi/VG-S-2,vgSr/VG-S-3,vgCh/VG-C-1,vgCi/VG-C-2,vgCr/VG-C-3}{%
  \draw[stn] (\c) circle (10mm);
  \node[inner sep=0] at (\c) {\includegraphics[height=9.5mm]{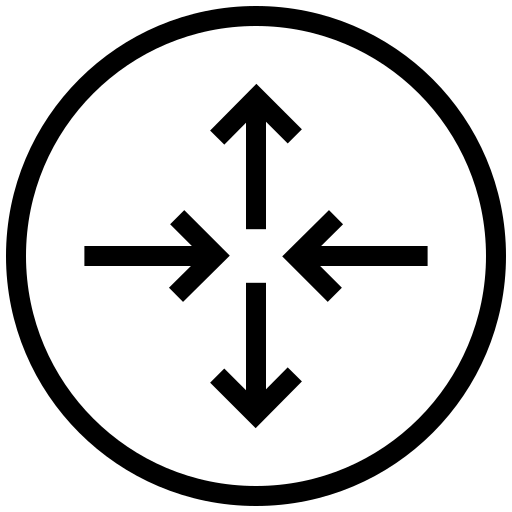}};
  \node[rlab] at ($(\c)+(0,-1.28)$) {\l};}
% fourth L2 vanguard in each set, retained but unused by any circuit (drawn faded)
\foreach \c/\l in {vgS4/VG-S-4,vgC4/VG-C-4}{%
  \draw[stn,draw=gray!35,dash pattern=on 1.2pt off 1.2pt] (\c) circle (10mm);
  \node[inner sep=0,opacity=0.45] at (\c) {\includegraphics[height=9.5mm]{tor_figure/router.png}};
  \node[rlab,text=gray!60] at ($(\c)+(0,-1.28)$) {\l};}
\foreach \c/\l in {mSi/M-S,mSr/M-S,mCi/M-C,mCr/M-C}{%
  \draw[stn] (\c) circle (11.4mm);
  \node[inner sep=0] at (\c) {\includegraphics[height=11.6mm]{tor_figure/router.png}};
  \node[rlab] at ($(\c)+(0,-1.34)$) {\l};}
% descriptor-circuit middle relays: label above, so it clears the line below them
\foreach \c/\l in {mSd/M-S,mCd/M-C}{%
  \draw[stn] (\c) circle (11.4mm);
  \node[inner sep=0] at (\c) {\includegraphics[height=11.6mm]{tor_figure/router.png}};
  \node[rlab] at ($(\c)+(0,1.34)$) {\l};}
% single entry guard per side: larger, because all three circuits share it
\foreach \c/\l in {egS/EG-S,egC/EG-C}{%
  \draw[stn,draw=gray!55,line width=0.9pt] (\c) circle (14mm);
  \node[inner sep=0] at (\c) {\includegraphics[height=13mm]{tor_figure/router.png}};
  \node[rlab] at ($(\c)+(0,-1.78)$) {\l};}
\draw[stn] (ip) circle (10.4mm);
\node[inner sep=0] at (ip) {\includegraphics[height=11.6mm]{tor_figure/router.png}};
\node[rlab] at ($(ip)+(0,1.3)$) {IntP};
\draw[stn] (rp) circle (10.4mm);
\node[inner sep=0] at (rp) {\includegraphics[height=11.6mm]{tor_figure/router.png}};
\node[rlab] at ($(rp)+(0,-1.34)$) {RP};
\fill[white] (hs) circle (10.5mm);
\node[inner sep=0] at (hs) {\includegraphics[height=20mm]{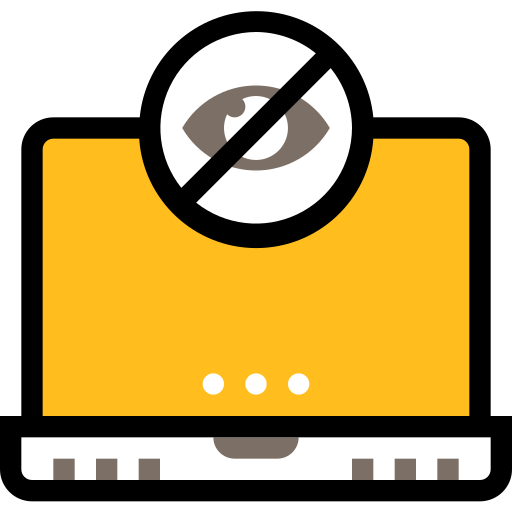}};
\node[elab] at ($(hs)+(0,-1.6)$) {Onion\\Service};
\fill[white] (cl) circle (7mm);
\node[inner sep=0] at (cl) {\includegraphics[height=22mm]{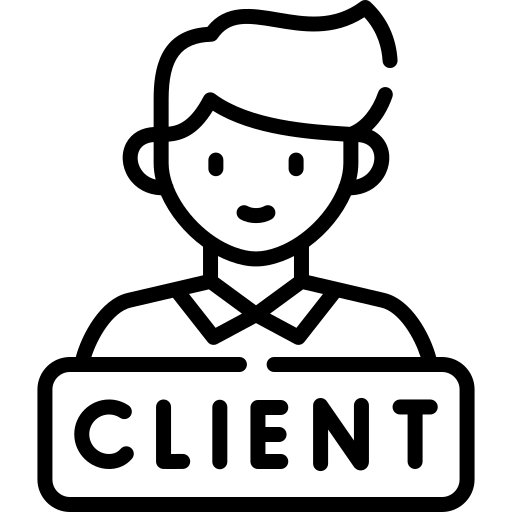}};
\node[elab] at ($(cl)+(0,-1.6)$) {Client};
\fill[white] (hd) circle (7.0mm);
\node[inner sep=0] at (hd) {\includegraphics[height=12mm]{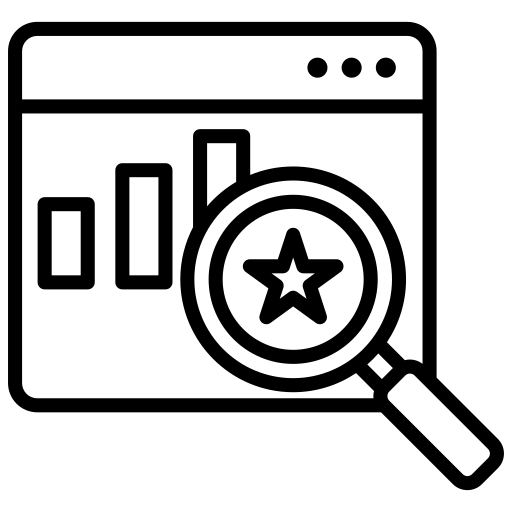}};
\node[elab] at ($(hd)+(0,1.10)$) {HSDir};
\end{tikzpicture}%
}
\caption{On each side, the
larger node is the shared entry guard and the faded node an unused layer-2
vanguard. Colour identifies the step and stroke its role (solid: request,
dashed: forwarded, dotted: acknowledgment); the double-headed arrow
(step~\textbf{9}) carries application data. Request--acknowledgment pairs take
suffixes \textbf{a}/\textbf{b} (\textbf{1a}/\textbf{1b}).}\label{fig:protocol-percircuit}
\end{figure*}

In an onion service protocol, the onion service first picks a few relays,
three by default, as its \emph{Introduction Points} (IntPs). It then builds an
\emph{introduction circuit} to each. Over this circuit, the service asks the
IntP to accept introductions for it (\texttt{ESTABLISH\_INTRO}, \textbf{1a}),
and the IntP confirms (\texttt{INTRO\_ESTABLISHED}, \textbf{1b}). The service
then publishes a signed \emph{descriptor} that lists its IntPs. It uploads the
descriptor to a few directory relays called \emph{HSDirs}, which confirm
storage (\textbf{2a}/\textbf{2b}). Which HSDirs hold the descriptor is
computed from the service's public key and the current time~\cite{10.1145/3589334.3645591}.
A client that knows the onion address can therefore find and download the
descriptor (\textbf{3a}/\textbf{3b}) and learn the service's IntPs.

To connect, the client picks a relay as a \emph{Rendezvous Point} (RP), where
it will later meet the service. It builds a circuit to the RP and registers
there (\texttt{ESTABLISH\_RENDEZVOUS}, \textbf{4a};
\texttt{RENDEZVOUS\_ESTABLISHED}, \textbf{4b}). Next, it builds a separate
circuit to one of the IntPs and sends an introduction request,
\texttt{INTRODUCE1} (\textbf{5a}). This request names the RP and carries key
material. The IntP acknowledges the request to the client
(\texttt{INTRODUCE\_ACK}, \textbf{5b}) and forwards it to the service as
\texttt{INTRODUCE2} over the introduction circuit (\textbf{6}). The service
then builds its own circuit to the RP and sends \texttt{RENDEZVOUS1}
(\textbf{7}). The RP joins the two circuits and passes the reply to the
client as \texttt{RENDEZVOUS2} (\textbf{8}). From then on, the client and
service exchange data through the RP (\textbf{9}). Figure~\ref{fig:protocol-percircuit}
summarizes each step of a connection.

\section{Threat Model and Attack Approach}
\label{sec:attack}
This section first defines the adversary precisely and then formalizes the set
of circuits among which the introduction circuit of the hidden service is
anonymous. It next describes the attack, explaining how intersecting repeated
observations can isolate a relay's successor and how this step is applied
adaptively to reconstruct the introduction circuit hop by hop
(Algorithm~\ref{alg:reconstruction}). It concludes with the assumptions under
which reconstruction succeeds.

\subsection{Threat Model}

\label{sec:tor-threat-model}

\noindent
\paragraph{\textbf{Tor threat model.}}
Tor does not aim to defend against a global adversary that observes the
entire network. Instead, its threat model considers a partial, potentially
active adversary that observes only a fraction of network traffic and may
inject traffic into the network. Such an adversary may control a subset of
Tor relays or observe traffic at network infrastructure, such as Internet
service providers (ISPs), autonomous systems (ASes), or Internet exchange
points (IXPs). Tor aims to prevent \emph{traffic analysis}, in which an adversary infers
who communicates with whom from observed traffic patterns, even when it
is a partial observer of the network with limited prior suspicion about the
parties involved~\cite{dingledine2004tor,murdoch2005low}.

Identifying a particular circuit through traffic analysis is challenging
because each relay multiplexes traffic from many concurrent circuits.
Moreover, ordinary Tor circuits are typically used for about $10$~min
before being replaced, limiting the evidence an adversary can accumulate
about a fixed path connecting two endpoints. Onion services present a
different case: an onion service's introduction circuits can remain active
for $18$--$24$~h~\cite{platzer2020critical}, and every introduction
handshake sent through a given Introduction Point traverses the same
service-side path.

\noindent
\paragraph{\textbf{Our adversary model.}}
The adversary's objective is to identify the network location of a target
onion service. The adversary (i) is \emph{adaptive} and initially knows only
the target's onion address; (ii) controls a Tor client from which it can send
ordinary introduction requests; and (iii) observes at most one
Tor relay at a time. The first relay it observes is an Introduction Point
selected from the service's descriptor. At each observed relay,
the adversary can obtain timestamped packet headers for traffic sent by
that relay. The adversary does not delay, drop, or modify relay traffic.
Its observations are passive, and its only active interaction with the
network consists of sending ordinary introduction requests from its client.

Apart from the onion address and the Introduction Point advertised in the
descriptor, the adversary has no prior knowledge of the service's network
location or the relays connecting the service to that Introduction Point.
It moves its observation point toward the service by using the result of
each stage to decide which relay to observe next.

\subsection{Attack Strategy}
\label{sec:attack-overview}

To identify the onion service's network location, the adversary first
retrieves the service's descriptor using its onion address, selects an
advertised Introduction Point, and obtains network-level visibility of
that relay. The adversary then repeatedly
triggers introduction handshakes from its client and records the destination
addresses of packets sent by the monitored relay during each handshake
(Section~\ref{sec:anonymity-set}). By intersecting the resulting anonymity
sets, the adversary narrows the candidates to a single relay, identifying
the monitored relay's successor toward the service on the introduction
circuit (Section~\ref{sec:intersection-attacks}).

The adversary then shifts its observation to the identified relay and
repeats the procedure, reconstructing the introduction circuit hop by hop
until it reaches the service's entry guard. From there, it seeks to identify
the service's network location.

\noindent
\paragraph{\textbf{Stages, iterations, and direction.}}
The attack proceeds in \emph{stages} $i=0,\ldots,K-1$. At stage~$i$, the adversary monitors relay $r_{m_i}$, where $r_{m_0}$ is the targeted Introduction Point selected from the descriptor, and seeks to identify its \emph{successor}. Each stage consists of \emph{iterations} $j=1,2,\ldots$, each corresponding to one introduction handshake triggered by the adversarial client.
Because reconstruction starts at the Introduction Point, we order the introduction circuit from the Introduction Point toward the service (Section~\ref{sec:onion-service}). Accordingly, the successor of $r_{m_i}$ is its neighbor one hop closer to the service, while its \emph{predecessor} is its neighbor one hop closer to the Introduction Point. For $i\ge 1$, the predecessor is the previously monitored relay $r_{m_{i-1}}$.

\noindent
\paragraph{\textbf{Target path.}}
An onion service's introduction circuit consists of four relays, the last
of which is the Introduction Point. Under the default Vanguards-Lite
configuration, the path is service $\to$ entry guard $\to$ layer-2
vanguard $\to$ middle relay $\to$ Introduction Point. Under the optional
Full Vanguards configuration, a layer-3 vanguard replaces the middle
relay, leaving the circuit length unchanged.\footnote{See
\href{https://spec.torproject.org/vanguards-spec/}{Tor Specifications,
``Tor Vanguards Specification.''}}
Reconstruction therefore requires $K=4$ stages in both configurations:
$r_{m_0},\ldots,r_{m_3}$ denote the Introduction Point, the middle relay
or layer-3 vanguard, the layer-2 vanguard, and the entry guard,
respectively. The successor of the entry guard is the onion-service
host itself, whose network address we denote by $\mathrm{HS}$.
Identifying this address completes the attack.

\subsection{Observation Windows and Successor Anonymity Sets}

\label{sec:anonymity-set}

For each iteration $j$ of stage $i$, we define the \emph{observation window}
$W_i^{(j)}$ as the interval between the adversarial client sending
\texttt{INTRODUCE1} and receiving the corresponding \texttt{RENDEZVOUS2}.
Every relay on the target introduction circuit must forward the resulting
\texttt{INTRODUCE2} cell (step~\textbf{6} in
Fig.~\ref{fig:protocol-percircuit}) within this interval. The cell is
generated only after the client sends \texttt{INTRODUCE1}, and the service
must receive it before sending \texttt{RENDEZVOUS1}, which precedes
\texttt{RENDEZVOUS2}.

For each window, the adversary constructs an \emph{anonymity
set}~\cite{diaz2002towards} containing the candidate addresses that cannot
yet be distinguished from the address of the successor of $r_{m_i}$
based on the available observations:
\[
\mathcal{A}_i^{(j)}
=
\left\{
a \;\middle|\;
\begin{array}{l}
a \text{ is the destination IP address of a packet}\\
\text{sent by } r_{m_i} \text{ during } W_i^{(j)}
\end{array}
\right\}.
\]
Here, $\operatorname{addr}(r)$ denotes the IP address of relay $r$.
Because the cell carrying the introduction must pass from $r_{m_i}$
to its successor within $W_i^{(j)}$, the set
$\mathcal{A}_i^{(j)}$ contains the successor's address,
$\operatorname{addr}(r_{m_{i+1}})$, or $\mathrm{HS}$ at the final stage,
provided that the target introduction circuit remains unchanged and
the packet timestamps are aligned with the client's clock. Every
other element is a candidate the adversary cannot yet rule out.

\noindent
\paragraph{\textbf{Predecessor exclusion.}}
\label{sec:predecessor-exclusion}
From stage~$1$ onward, $r_{m_i}$ receives the introduction cell from its
predecessor $r_{m_{i-1}}$. Transport-level replies, such as TCP
acknowledgments, can therefore cause $\operatorname{addr}(r_{m_{i-1}})$
to recur in every observation window and prevent the anonymity set from
converging to a single address. The adversary addresses this by
intersecting the filtered sets
\[
\tilde{\mathcal{A}}_i^{(j)} =
\begin{cases}
    \mathcal{A}_i^{(j)}, & i = 0,\\
    \mathcal{A}_i^{(j)} \setminus
    \{\operatorname{addr}(r_{m_{i-1}})\}, & i \ge 1.
\end{cases}
\]

This filtering removes the true successor only if it shares an IP address
with $r_{m_{i-1}}$. Tor's general rule against placing two relays from
the same \texttt{/16} subnet on a circuit does not exclude this case:
Vanguards-Lite relaxes subnet and family restrictions for onion-service
circuits.\footnote{See
\href{https://spec.torproject.org/vanguards-spec/path-construction.html}{Tor
Specifications, ``Vanguard Path Construction.''}}
We therefore assume that each successor's address differs from its
predecessor's (Section~\ref{sec:progressive-reconstruction}).
At stage~$0$, the Introduction Point also sends traffic, such as
\texttt{INTRODUCE\_ACK}, to the last hop of the adversary's client-side
introduction circuit. To prevent this hop from recurring in every window,
the adversary builds a new client-side circuit for each iteration
(in our experiments, by starting a fresh Tor client;
Section~\ref{sec:experimental-setup}). Consequently, this hop generally
varies across windows and is eliminated, like other varying destinations,
by the intersection attack described in the next subsection.

\noindent
\paragraph{\textbf{Operational construction.}}
The adversary can obtain timestamped packet headers for $r_{m_i}$
through ISP cooperation or by compromising the relay itself. Independently,
the adversary timestamps the \texttt{INTRODUCE1} and
\texttt{RENDEZVOUS2} events at its client. It then collects the
destination IP addresses from outgoing packet headers whose timestamps
fall within $W_i^{(j)}$. This construction requires only temporal
correlation between the client-side handshake events and the packet
timestamps at the monitored relay, assuming that the client and capture
clocks are synchronized to an accuracy well within the duration of
$W_i^{(j)}$.

\subsection{Intersection Attack}

\label{sec:intersection-attacks}

An \emph{intersection attack}~\cite{danezis2004statistical} reduces an
anonymity set by intersecting observations from different windows. It
succeeds when the target persists across observations while unrelated
candidates vary. Because $r_{m_i}$ concurrently carries unrelated traffic,
a single $\tilde{\mathcal{A}}_i^{(j)}$ may be large. The adversary therefore
intersects the sets from successive iterations:
$\mathcal{I}_i^{(1)}=\tilde{\mathcal{A}}_i^{(1)}, \qquad
\mathcal{I}_i^{(j)}=\mathcal{I}_i^{(j-1)}\cap\tilde{\mathcal{A}}_i^{(j)}
\quad (j>1).$

Under the conditions of Section~\ref{sec:anonymity-set}, the true successor
is contained in every $\tilde{\mathcal{A}}_i^{(j)}$ and is therefore
preserved by intersection. Stage~$i$ converges at the first iteration $j$
for which $|\mathcal{I}_i^{(j)}|=1$. At that point, the remaining address
is identified as the successor of $r_{m_i}$. An empty intersection
indicates that the true successor was missing from at least one window.
Algorithm~\ref{alg:reconstruction} treats this as the target introduction
circuit having been dropped.

\noindent
\paragraph{\textbf{Dynamic-traffic assumption.}}

Convergence requires that, at each stage~$i$, every candidate in
$\tilde{\mathcal{A}}_i^{(1)}$ other than the true successor is absent
from at least one later window before the target introduction circuit
is dropped.

Each window spans a single handshake, lasting approximately one second
(Appendix~\ref{app:introduction-latency}). An unrelated address therefore
survives only if $r_{m_i}$ sends a packet to it in \emph{every} window.
We expect this to be rare: unrelated circuits do not transmit
continuously, new connections may use fresh circuits over different
relays after approximately $10$~min in ordinary Tor, and Tor's
\emph{client} population turns over approximately $2.5$ times per
day~\cite{jansen2016safely}.

An unrelated address that appears in every window through iteration $j$
necessarily remains in $\mathcal{I}_i^{(j)}$ and delays convergence.
It prevents convergence only if it continues to appear until the target
introduction circuit is dropped. We examine such persistent candidates
empirically in Section~\ref{sec:rq2}.

\subsection{Progressive Circuit Reconstruction}

\label{sec:progressive-reconstruction}

\noindent
\paragraph{\textbf{Stage progression.}}
Reconstruction begins at the advertised Introduction Point $r_{m_0}$.
At each stage~$i$, the adversary obtains network visibility at
$r_{m_i}$, repeatedly triggers introduction handshakes, and intersects
the resulting anonymity sets
(Sections~\ref{sec:anonymity-set}--\ref{sec:intersection-attacks}).
If the intersection converges to a singleton, the remaining address
identifies the next monitored relay $r_{m_{i+1}}$. At the final stage,
where the monitored relay is the entry guard, the remaining address
instead identifies $\mathrm{HS}$.

Algorithm~\ref{alg:reconstruction} formalizes the complete procedure.
It aborts with \textsc{IntroductionCircuitDropped} if an intersection
becomes empty, rather than selecting a spurious relay.

\begin{algorithm}[t]

\caption{Progressive Introduction-Circuit Reconstruction}

\label{alg:reconstruction}

\begin{algorithmic}[1]

\Require Introduction Point $r_{m_0}$; number of stages $K$
($K=4$ for a service-side introduction circuit)

\Ensure Reconstructed path
$(r_{m_0},\ldots,r_{m_{K-1}},\mathrm{HS})$ or
\textsc{IntroductionCircuitDropped}

\For{stage $i \gets 0$ to $K-1$}

    \State Obtain network visibility of $r_{m_i}$

    \State $j \gets 0$

    \Repeat

        \State $j \gets j+1$

        \State Trigger an introduction handshake through $r_{m_0}$
        over a fresh client-side circuit

        \State Observe $\mathcal{A}_i^{(j)}$ at $r_{m_i}$

        \State Compute $\tilde{\mathcal{A}}_i^{(j)}$
        \Comment{exclude $\operatorname{addr}(r_{m_{i-1}})$ if $i>0$}

        \State $\mathcal{I}_i \gets
        \bigcap_{\ell=1}^{j}\tilde{\mathcal{A}}_i^{(\ell)}$
        \Comment{running value of $\mathcal{I}_i^{(j)}$}

        \If{$|\mathcal{I}_i| = 0$}
            \Comment{true successor missing from a window}

            \State \Return \textsc{IntroductionCircuitDropped}

        \EndIf

    \Until{$|\mathcal{I}_i| = 1$}

    \If{$i = K-1$}

        \State $\mathrm{HS} \gets$ unique element of $\mathcal{I}_i$

    \Else

        \State $r_{m_{i+1}} \gets$ relay at the unique address in
        $\mathcal{I}_i$

    \EndIf

\EndFor

\State \Return $(r_{m_0},\ldots,r_{m_{K-1}},\mathrm{HS})$

\end{algorithmic}

\end{algorithm}

\noindent
\paragraph{\textbf{Assumptions and limitations.}}
Beyond the dynamic-traffic assumption (Section~\ref{sec:intersection-attacks}),
reconstruction relies on three assumptions. First, the target introduction
circuit must remain unchanged from the first iteration of stage~$0$ until
stage~$K-1$ converges. The algorithm imposes no iteration limit, but the
circuit's lifetime bounds each attempt in practice. Second, packet
timestamps at the monitored relay must be aligned with the client's
handshake timestamps (Section~\ref{sec:anonymity-set}). Misalignment may
cause the true successor to be excluded from $\mathcal{A}_i^{(j)}$.
Third, predecessor exclusion assumes that the successor and predecessor
of each monitored relay have different IP addresses
(Section~\ref{sec:predecessor-exclusion}). If they share an address,
the filtering step removes the true successor, preventing the stage
from identifying it.

\section{Evaluation}

\label{sec:eval}

In this section, we seek to answer one main question: \emph{can the attack
reconstruct the introduction circuit of a live onion service, under
third-party background traffic and within the circuit's lifetime?}
To answer it, we ran Algorithm~\ref{alg:reconstruction} nine times on the
live Tor network against an onion service that we operate, and estimated
the reconstruction time from the number of iterations each stage needed to
isolate the target.

The adversary in our threat model observes a relay's traffic from the
network infrastructure it is connected to, such as its ISP or an upstream
exchange point, without running the relay itself. We cannot compromise
such infrastructure, so we use self-operated relays that we operate as a proxy for that
vantage point: we pinned the service's introduction circuit to four public
Tor relays of ours, and the attack observed traffic only at those relays.
The relays had been active for $69$--$126$ days when the experiments began
and were eligible as entry guards and middle relays, so they carried genuine
traffic from other Tor users. This gave us realistic background traffic
without instrumenting third-party infrastructure or targeting third-party
circuits, and, through the pinned path, the ground truth for every stage. We discuss how faithfully this setup reflects the threat model in
Appendix~\ref{app:implementation}.

\subsection{Experimental Setup}

\label{sec:experimental-setup}

\noindent
\paragraph{\textbf{Metrics.}}
A primary metric is the effort required to carry out the attack, which we
measure in terms of the number of iterations needed for each stage to
converge. For every iteration $j=1,2,\ldots$ of stage~$i$, we record the
cardinality $|\mathcal{I}_i^{(j)}|$ of the running intersection, so each
stage yields a trajectory
$(|\mathcal{I}_i^{(1)}|,|\mathcal{I}_i^{(2)}|,\ldots)$. From this trajectory,
we derive the initial candidate-set size
$|\mathcal{I}_i^{(1)}|=|\tilde{\mathcal{A}}_i^{(1)}|$ (after predecessor
exclusion for $i\geq1$), the convergence iteration
$T_{\mathrm{conv}}=\min\{j:|\mathcal{I}_i^{(j)}|=1\}$, and the threshold
crossings $T_{\leq q}=\min\{j:|\mathcal{I}_i^{(j)}|\leq q\}$ for
$q\in\{10,3,2\}$.

Wall-clock time depends on the prototype implementation, so we treat it as
a secondary metric and derive it from the iteration count below. At the
start of each stage, we also record the monitored relay's
CW as reported by Onionoo.\footnote{Onionoo is a Tor Project service that
provides current information about Tor relays and bridges, including relay
status, bandwidth, and consensus-weight data.}

\noindent
\paragraph{\textbf{Observation procedure.}}
The monitored relay runs four processes: (i)~a capture process that streams
its outgoing packet headers, (ii)~the intersection plugin that consumes
them, (iii)~the colocated probe client that bounds each observation window,
and (iv)~a controller that drives the loop. The capture reaches the plugin
through a pipe.\footnote{A pipe is a kernel-managed buffer for
unidirectional inter-process communication; its contents never reach disk.}
The plugin maintains the running intersection, admitting a destination
address into $\mathcal{A}_i^{(j)}$ only between two signals that delimit the
observation window defined in Section~\ref{sec:anonymity-set}. The probe
client raises these signals immediately before sending
\texttt{INTRODUCE1} and upon receiving the corresponding
\texttt{RENDEZVOUS2}. Because the client and relay share a host, these
signals bound each window without requiring timestamp synchronization.

The controller relaunches the client after $\delta=30$~s and checks whether
the intersection has converged to a single candidate. Once it has, the
procedure is repeated at the identified successor, which is another relay
we operate. No capture file is written; addresses, keys, and intersections
remain only in volatile memory. Colocation provides only the observation
window boundaries and simulates observing a network infrastructure. Network visibility comes from operating the relays
ourselves, and the arrangement does not otherwise relax the threat model,
as Appendix~\ref{app:implementation} explains alongside the step-by-step
procedure corresponding to Figure~\ref{fig:protocol-colocated}.

Pinning the target circuit through these relays places every hop at a
location where we have network-level visibility and establishes the ground
truth for each stage. This ground truth is used only to verify the result,
never to compute it.

\subsection{Results}

\label{sec:rq1}

\noindent
\paragraph{\textbf{End-to-end reconstruction.}}

Across the nine end-to-end experiments, the attack successfully
reconstructed the complete introduction path in every run. All $36$ stages
converged to the correct successor. A complete reconstruction required a
median of $260$ iterations, ranging from $101$ to $770$ iterations across
runs (Table~\ref{tab:end_to_end}).

Converting these iteration counts into wall-clock time requires assumptions
about both the cost of each iteration and the time required to obtain
visibility at each monitored relay. We model the reconstruction time of a
run as
\begin{equation}
\label{eq:cost}
T=(31\,\mathrm{s})N+4v,
\end{equation}
where $N$ is the total number of iterations across the four stages and $v$
is the time required to obtain visibility at one monitored relay, incurred
once per stage. The $31$~s iteration cost consists of the $\delta=30$~s
delay between iterations in our experimental setup and approximately $1$~s
for the introduction handshake itself. The $30$~s delay is a configurable
parameter of our implementation and could be reduced by an adversary. The
approximately $1$~s handshake duration is based on measurements of public
onion services reported in Appendix~\ref{app:introduction-latency}.

We do not include the Tor client bootstrap time incurred by our prototype
in Eq.~\eqref{eq:cost} because it is an implementation cost rather than an
intrinsic requirement of the attack.\footnote{Our prototype launches a fresh client for every iteration, which must fetch the consensus and the service descriptor before sending \texttt{INTRODUCE1}; an adversary can cache both.} Visibility in our experiments came from operating the relays
ourselves, so the experiments provide no empirical estimate of $v$. We
therefore leave $v$ as a free parameter and report reconstruction cost as a
function of it.

As shown in Table~\ref{tab:end_to_end}, when $v=0$, the median
reconstruction takes $2.24$~h and the slowest takes $6.63$~h. These
correspond to approximately $12\%$ and $37\%$, respectively, of $18$~h,
the lower end of the introduction circuit's $18$--$24$~h lifetime
(Section~\ref{sec:tor-threat-model}). Table~\ref{tab:end_to_end} also shows
how $T$ increases with $v$ and reports each run's break-even visibility
budget $v_{\max}$, defined as the largest value of $v$ for which the
reconstruction still completes within $18$~h. The median run permits
$v_{\max}=3.94$~h per relay, while the most iteration-intensive run,
experiment~6, permits $2.84$~h. At $v=4$~h, five of the nine runs exceed
the $18$~h bound.

The visibility cost becomes the dominant component of the modeled
reconstruction time well before this point. For the median run, obtaining
visibility accounts for approximately $64\%$ of the total time at
$v=1$~h and $88\%$ at $v=4$~h. Across all runs, $v_{\max}$ ranges from
$2.84$ to $4.28$~h despite a more than sevenfold difference in total
iteration counts.

\begin{table*}[t]

\centering

\scriptsize

\setlength{\tabcolsep}{2pt}

\caption{End-to-end reconstruction cost across the nine experiments: iterations
to convergence per stage (stage-start consensus weight in parentheses), their
sum $N$, and $T$ from Eq.~\eqref{eq:cost} for several per-relay visibility
costs $v$ (bold: exceeds $18$~h). $v_{\max}$ is the largest $v$ for which the
run still completes within $18$~h.}

\label{tab:end_to_end}

\begin{tabular}{l r r r r r r r r r}

\toprule

&
\multicolumn{4}{c}{Iterations per stage (CW)} & &

\multicolumn{3}{c}{$T$ in hours at cost $v$}
& \\

\cmidrule(lr){2-5}\cmidrule(lr){7-9}

\textbf{Run}
&

\textbf{IntP}
&

\textbf{Middle}
&

\textbf{Vanguard}
&

\textbf{Entry Guard}
&

\shortstack[r]{\textbf{Total}\\$N$}
&

$v{=}0$
& $v{=}1$~h & $v{=}4$~h & $v_{\max}$ \\

\midrule

1 & 285 (830) & 5 (5000) & 61 (1200) & 31 (9300) & 382 & 3.29 &
7.29 & \textbf{19.29} & 3.68 \\

2 & 52 (850) & 13 (5000) & 4 (1200) & 62 (9300) & 131 & 1.13 &
5.13 & 17.13 & 4.22 \\

3 & 163 (850) & 7 (4000) & 12 (1200) & 78 (9300) & 260 & 2.24 &
6.24 & \textbf{18.24} & 3.94 \\

4 & 8 (850) & 11 (4000) & 10 (1200) & 123 (9900) & 152 & 1.31 &
5.31 & 17.31 & 4.17 \\

5 & 178 (1500) & 17 (4100) & 142 (1200) & 167 (4800) & 504 & 4.34 &
8.34 & \textbf{20.34} & 3.42 \\

6 & 324 (1400) & 10 (4100) & 251 (1200) & 185 (4800) & 770 & 6.63 &
10.63 & \textbf{22.63} & 2.84 \\

7 & 11 (1900) & 10 (4800) & 24 (1300) & 166 (4800) & 211 & 1.82 &
5.82 & 17.82 & 4.05 \\

8 & 8 (1900) & 53 (4800) & 6 (1300) & 193 (4800) & 260 & 2.24 &
6.24 & \textbf{18.24} & 3.94 \\

9 & 10 (2000) & 4 (4800) & 3 (1300) & 84 (4400) & 101 & 0.87 &
4.87 & 16.87 & 4.28 \\

\midrule

\textbf{Median}
& 52 & 10 & 12 & 123 & 260 & 2.24 & 6.24 &
\textbf{18.24} & 3.94 \\

\bottomrule

\end{tabular}

\end{table*}

\begin{table*}[t]
\centering
\scriptsize
\setlength{\tabcolsep}{1.8pt}
\caption{Representative within-stage convergence. $|\tilde{\mathcal{A}}_i^{(1)}|$
is the initial candidate-set size after predecessor exclusion, $T_{\leq k}$ the
first iteration with at most $k$ candidates, $T_{\mathrm{conv}}$ the iteration
of convergence to a singleton, and CW the monitored relay's stage-start
consensus weight. Complete results: Appendix~\ref{app:run-table}.}
\label{tab:stage-contrasts}
\resizebox{\linewidth}{!}{%
\begin{tabular}{@{}lrrrrrr@{\hspace{7pt}}lrrrrrr@{}}
\toprule
\textbf{Run, day, time} & \textbf{CW} & $|\tilde{\mathcal{A}}_i^{(1)}|$ & $T_{\leq10}$ & $T_{\leq3}$ & $T_{\leq2}$ & $T_{\mathrm{conv}}$ &
\textbf{Run, day, time} & \textbf{CW} & $|\tilde{\mathcal{A}}_i^{(1)}|$ & $T_{\leq10}$ & $T_{\leq3}$ & $T_{\leq2}$ & $T_{\mathrm{conv}}$ \\
\midrule
\multicolumn{7}{@{}l}{\textit{Introduction Point}} & \multicolumn{7}{l}{\textit{Vanguard}} \\
R4 D2 02:00 & 850  & 81  & 3 & 4   & 4   & 8   & R9 D4 10:00 & 1300 & 48  & 2  & 3  & 3  & 3   \\
R2 D1 10:00 & 850  & 101 & 2 & 9   & 52  & 52  & R3 D1 18:00 & 1200 & 85  & 2  & 4  & 12 & 12  \\
R1 D1 02:00 & 830  & 79  & 3 & 42  & 245 & 285 & R5 D2 10:00 & 1200 & 148 & 38 & 61 & 76 & 142 \\
R6 D2 18:00 & 1400 & 84  & 3 & 233 & 324 & 324 & R6 D2 18:00 & 1200 & 161 & 11 & 49 & 66 & 251 \\
\midrule
\multicolumn{7}{@{}l}{\textit{Middle}} & \multicolumn{7}{l}{\textit{Entry Guard}} \\
R9 D4 10:00 & 4800 & 90  & 2 & 3  & 3  & 4  & R1 D1 02:00 & 9300 & 218 & 8  & 16 & 31  & 31  \\
R6 D2 18:00 & 4100 & 147 & 3 & 9  & 10 & 10 & R4 D2 02:00 & 9900 & 256 & 13 & 45 & 49  & 123 \\
R5 D2 10:00 & 4100 & 158 & 6 & 13 & 16 & 17 & R6 D2 18:00 & 4800 & 98  & 7  & 23 & 177 & 185 \\
R8 D4 02:00 & 4800 & 198 & 7 & 14 & 14 & 53 & R8 D4 02:00 & 4800 & 191 & 16 & 62 & 62  & 193 \\
\bottomrule
\end{tabular}}
\end{table*}

\noindent
\paragraph{\textbf{Rapid elimination.}}
\label{sec:rq2}
Across the $36$ stages in our empirical evaluation, the median iteration at
which $|\mathcal{I}_i^{(j)}|\leq10$ was three, and $64\%$ of stages reached
this threshold within five iterations. Comparing per-stage medians, the
candidate set shrank by $72$--$87\%$ after the first intersection and by
$85$--$96\%$ by the third iteration, depending on the stage.

We attribute this rapid reduction primarily to two factors, although our
privacy-preserving measurements retain no traffic information that would
allow us to separate their effects. First, the observation windows are
short: the handshakes measured in
Appendix~\ref{app:introduction-latency} complete in $0.521$--$1.780$~s,
depending on the service. An unrelated circuit that does not transmit
continuously is therefore unlikely to send a packet in every observation
window. Second, circuit positions not constrained by persistent guard or
Vanguard sets are selected from a large population of eligible relays, so
the destinations observed by a monitored relay can change as unrelated
circuits expire and are rebuilt. Unrelated candidates therefore tend to
appear and disappear across iterations, while the true successor remains
fixed on the target introduction circuit. This behavior is consistent with
the dynamic-traffic assumption in
Section~\ref{sec:intersection-attacks}.

  \paragraph{Persistent tail.}A few candidates recur persistently across anonymity sets and therefore dominate
the remaining convergence cost. This variation occurs even among relays with
the same CW: three of the six Vanguard stages at CW $1200$
converged after $12$, $142$, and $251$ iterations, while two of the three Middle
stages at CW $4800$ required $4$ and $53$, respectively
(Table~\ref{tab:stage-contrasts}). Initial anonymity-set size is similarly not predictive of convergence time. For
example, the Introduction Point stages of Runs~1, 4, and~6 started with similar
sets of $79$--$84$ candidates and reached $|\mathcal{I}|\leq10$ within three
iterations, yet converged after $285$, $8$, and $324$ iterations, respectively.

 In $13$ of the $36$ stages the
  intersection still held three or more candidates at the penultimate iteration and
  then fell to a singleton in one step. In seven of these the cardinality had been
  constant for at least ten iterations beforehand. The vanguard stage of Run~1 held
  six candidates for $57$ iterations and converged at iteration~$61$. The
  Introduction Point stages of Runs~3 and~5 held three for $152$ and $148$
  iterations before converging at iterations~$163$ and~$178$. Independent
  unrelated destinations would drop out gradually.

  One plausible explanation for these trajectories is persistent third-party
traffic, although our measurements cannot confirm its source. Because we observe
the monitored relay's outgoing packets, an unrelated destination may recur when
the relay maintains a persistent connection with another node. For example, if
the monitored relay serves as an entry guard for another Tor user, and a long-lived
stream is sent, such as an SSH session or file transfer, may repeatedly produce traffic
toward that user across our observation windows. Multiple concurrent circuits
from the same user may similarly cause relays selected from persistent path
components, such as guards or Vanguards, to recur across observations
(Section~\ref{sec:path-selection}). Our measurements cannot distinguish these cases. The plugin pseudonymizes
destination addresses and retains only intersection cardinalities, so persistent
candidates cannot be associated with particular circuits, users, traffic
streams, or path-selection mechanisms with certainty.

\noindent
\paragraph{\textbf{Limitations.}}
Our evaluation has two main limitations. First, it covers only the range of
consensus weights represented by the relays we were able to operate during
the measurement period. Relays with substantially higher consensus weight
were difficult to obtain and maintain, so we do not establish how
convergence behaves at the highest relay weights.
Second, the attack implementation is deliberately unoptimized. For example,
an adversary could pause its own introduction requests and remove candidates
that continue to receive background traffic during the pause. This could
reduce the persistent tail of the intersection, but it also risks eliminating
the true successor if that relay itself carries persistent third-party
traffic.

Our experiments are therefore intended to show that the current
$18$--$24$~h lifetime of introduction circuits can provide sufficient time
for path reconstruction, rather than to characterize the fastest possible
attack. The underlying exposure arises from the persistence of the
service-side path. Rebuilding that internal path on a timescale comparable
to ordinary Tor circuits (Section~\ref{sec:mitigation}) removes the long-lived
path on which the attack relies.
\section{Mitigation: Short-Lived Introduction Circuits}
\label{sec:mitigation}

Our attack exploits the $18$--$24$~h introduction-circuit lifetime: every
handshake the adversary triggers crosses the same service-side path, and
evidence accumulates until that path is reconstructed. We propose bounding that
lifetime. The mitigation rests on one assumption, which we state rather than
establish: \emph{a service-side path used for only about $10$~min, the lifetime
of an ordinary Tor circuit, does not give the adversary enough time to
reconstruct it.}

\noindent
\paragraph{\textbf{Mechanism.}}
The onion service retains its Introduction Point while
periodically rebuilding only the internal path leading to that relay,
approximately every $10$~min. Relay selection remains unchanged: entry guards
are still selected from the persistent guard set, and the vanguard layers
follow their existing selection rules (Section~\ref{sec:path-selection}).
Because the service descriptor identifies the Introduction Point but not the
relays behind it (Section~\ref{sec:onion-service}), rebuilding the internal
path does not require a descriptor update.
\noindent \paragraph{\textbf{Effect on the attack.}}
Only observations collected within a single rebuild interval correspond to the
same path. Once the internal path changes, observations accumulated against the
previous path no longer narrow the new one, effectively resetting the
path-specific evidence used by our attack. The rebuild interval therefore
limits not only the time available for intersection at each stage, but also the
time available to transition between stages. Identifying the successor of a
monitored relay does not automatically provide visibility at that successor;
the adversary must first obtain or coordinate the required network visibility
before continuing the attack. Thus, even if a stage converges within a
$10$-minute interval, the remaining time for obtaining visibility at the next
relay and completing subsequent stages is limited.

\noindent\paragraph{\textbf{Limits.}}
The mitigation is only as strong as the assumption it rests on. It gives an
introduction circuit the path lifetime of a rendezvous circuit, so it protects
the service only if rendezvous circuits themselves resist intersection attacks
of this kind.

\begin{comment}
\section{Related Work}
Attacks against Tor can be broadly categorized as denial-of-service (DoS) or
deanonymization attacks~\cite{iacovazzi2019duster}. We focus on the latter,
which can be further classified as \emph{passive}, where the adversary only
observes network traffic, or \emph{active}, where the adversary also manipulates
traffic \cite{karunanayake2021anonymisation}.
\end{comment}

\section{Related Work}

Attacks against Tor can be broadly categorized as either
denial-of-service (DoS) or deanonymization
attacks~\cite{iacovazzi2019duster}. This paper focuses on the latter,
which can be further classified as \emph{passive}, where the adversary
only observes network traffic, or \emph{active}, where the adversary
also manipulates traffic~\cite{karunanayake2021anonymisation}.

\label{sec:related}
\noindent \paragraph{\textbf{Passive attacks.}}One extensively studied deanonymization technique is website fingerprinting,
which infers a Tor client's destination from encrypted-traffic features such as
timing, direction, packet counts, and burst patterns~\cite{karunanayake2021anonymisation}.
Hayes and Danezis~\cite{hayes2016k} propose \emph{k-fingerprinting}, which uses
random forests to classify traffic against fingerprints of previously observed
websites. Such attacks require a prior suspicion of candidate websites,  and corresponding training
traces and, unlike ours, target client browsing rather than an onion service's
network location. Kwon et al.~\cite{kwon2015circuit} combine \emph{circuit fingerprinting} with website fingerprinting to deanonymize onion services, where circuit-level features first isolate service-side activity, after which website fingerprinting associates the trace with a monitored service. The attack requires prior suspicion to collect training traces and visibility near the service; the circuit-level information it needs may be available to an ISP under some conditions but is obtained more effectively by controlling the service's entry guard. DeepCorr~\cite{nasr2018deepcorr} instead learns to correlate
traffic observed at two network vantage points and generalizes to unseen flows
without destination-specific training. However, under partial network
observation, applying such correlation still requires prior suspicion to where to
observe both ends of the target flow. Our attack instead begins from a public
Introduction Point and determines the next observation point sequentially.

\noindent \paragraph{\textbf{Active attacks.}} 
Øverlier and Syverson introduced an attack that required no prior knowledge of the service's network location ~\cite{overlier2006locating}. Their attack operates Tor relays and repeatedly induces circuit construction until an adversarial relay is selected adjacent to the service, where traffic confirmation reveals its IP address. Persistent entry guards were introduced to mitigate this repeated-placement strategy~\cite{karunanayake2021anonymisation}, and as a result can not be used for deanonymization with no prior suspicion of a special target. Another representative example is DUSTER \cite{iacovazzi2019duster},
which embeds watermarks by exploiting Tor’s congestion control to correlate
onion services with their IP addresses. The watermark is introduced into the
circuit by the attacker and later detected at an adversarial controlled entry relay, allowing the attacker to
link the service to its underlying IP. However, controlling a trageted hidden service's entry node could be challenging.
\section{Conclusion}
\label{sec:conclusion}

\begin{comment}
The long lifetime of onion-service introduction circuits enables a
protocol-driven intersection attack that requires limited prior knowledge of the
service's network location. By repeatedly triggering introduction handshakes,
the adversary can exploit recurring traffic patterns to determine where in the
network to observe next, while requiring visibility of only one relay at a time.
We evaluated the attack in nine end-to-end experiments against a self-operated
onion service whose introduction circuit traversed our public Tor relays while
they concurrently carried genuine third-party traffic. In all nine experiments,
the attack successfully reconstructed the complete introduction path and
revealed the service's network location. We therefore proposed shortening the lifetime of the introduction circuit's
internal path while keeping its Introduction Point fixed. Periodically rebuilding
the internal path limits the time available to accumulate evidence against a
fixed circuit and to progress through its successive relays, while avoiding the
descriptor updates that would be required if the Introduction Point itself were
frequently replaced.
\end{comment}

This paper demonstrated that the long lifetime of onion-service introduction
circuits enables a protocol-driven intersection attack that requires only
limited prior knowledge of the service's network location. We presented an
attack in which an adaptive adversary with limited network visibility
repeatedly triggers introduction handshakes and exploits recurring traffic
patterns to determine where to observe next, requiring visibility of only
one relay at a time.
We evaluated the attack in nine end-to-end experiments against a
self-operated onion service whose introduction circuit traversed public Tor
relays that we operated while they concurrently carried genuine third-party
traffic. Across all nine experiments, the attack successfully reconstructed
the complete introduction path and identified the service's network location.
To mitigate this risk, we propose shortening the lifetime of the internal
path of the introduction circuit while keeping its Introduction Point fixed.
This removes the long-lived path on which the attack depends without
requiring frequent replacement of the Introduction Point.

\noindent
%\textbf{Artifacts and ethics.}
%The modified Tor daemon, the intersection plugin, the scripts that regenerate
%every table and figure, and step-by-step instructions for repeating the
%experiments with our source code will be made publicly available after the
%review process.
%The ethical considerations of the live-network experiments, structured along
%the Tor Research Safety Board's principles, are given in
%Appendix~\ref{sec:ethics}.
\bibliographystyle{splncs04}
\bibliography{refs}

% References do NOT count toward the page limit.

% Appendices (committee not obligated to review; keep the body self-contained).
 \appendix
 %\input{sections/appendix_onion_service_protocol}
 %\FloatBarrier
 \section{Per-Run and Per-Stage Convergence}
\label{app:run-table}

\begin{figure*}[tbp]
\centering
\includegraphics[width=\textwidth]{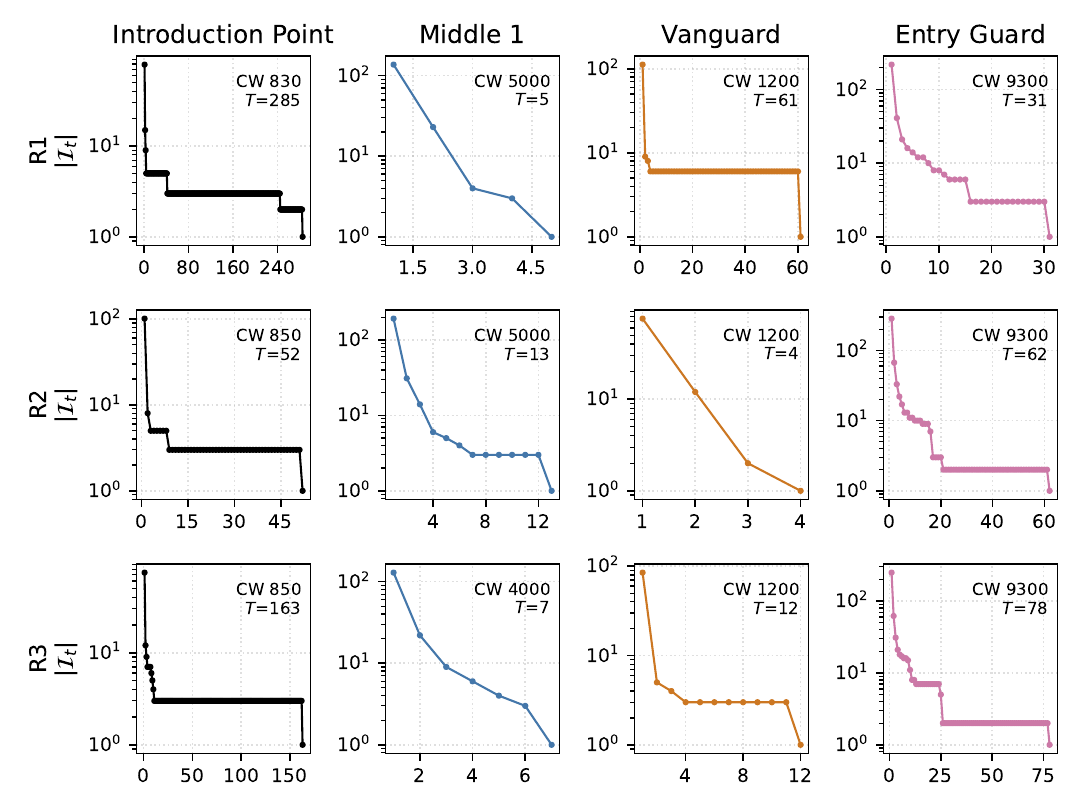}
\caption{Intersection degradation for Runs~1--3 across the four reconstruction
stages. The horizontal axis shows iteration $t$; the ordinal vertical axis
shows each distinct $|\mathcal{I}_t|$. Each panel reports stage-start consensus
weight (CW) and singleton convergence $T_{\mathrm{conv}}$.}
\label{fig:runs-grid-a}
\end{figure*}

\begin{figure*}[tbp]
\centering
\includegraphics[width=\textwidth]{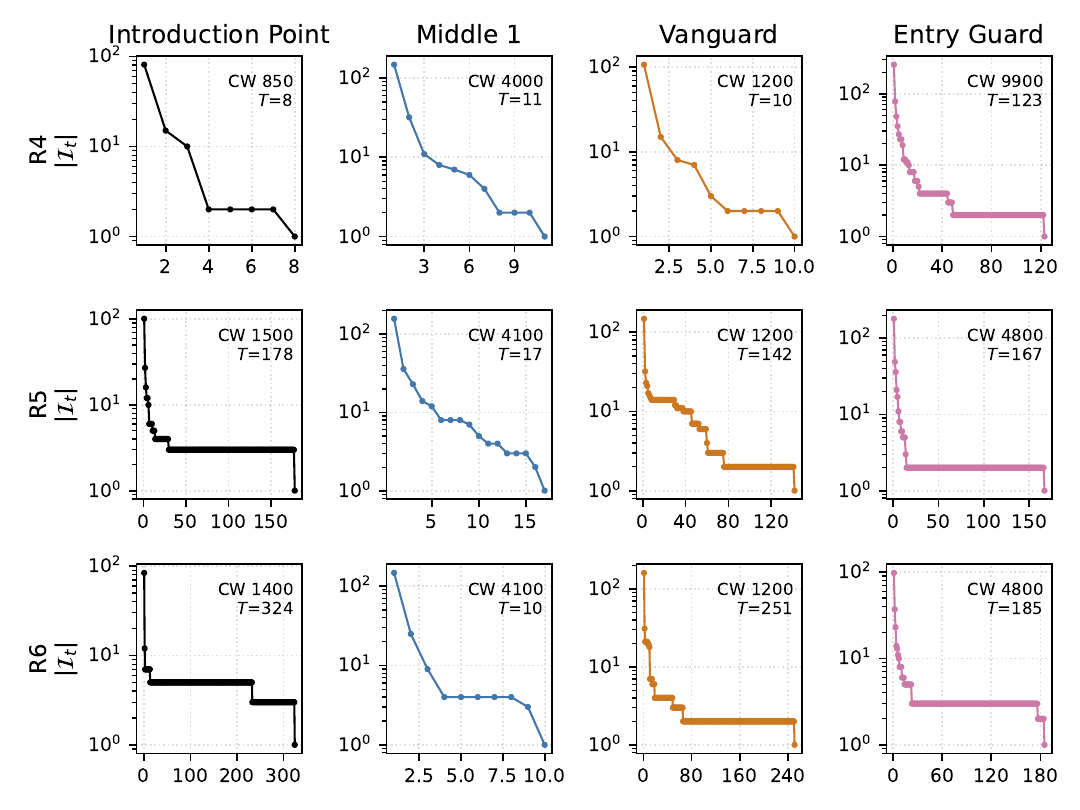}
\caption{Intersection degradation for Runs~4--6 across the four reconstruction
stages. Axes and annotations follow Figure~\ref{fig:runs-grid-a}.}
\label{fig:runs-grid-b}
\end{figure*}

\begin{figure*}[tbp]
\centering
\includegraphics[width=\textwidth]{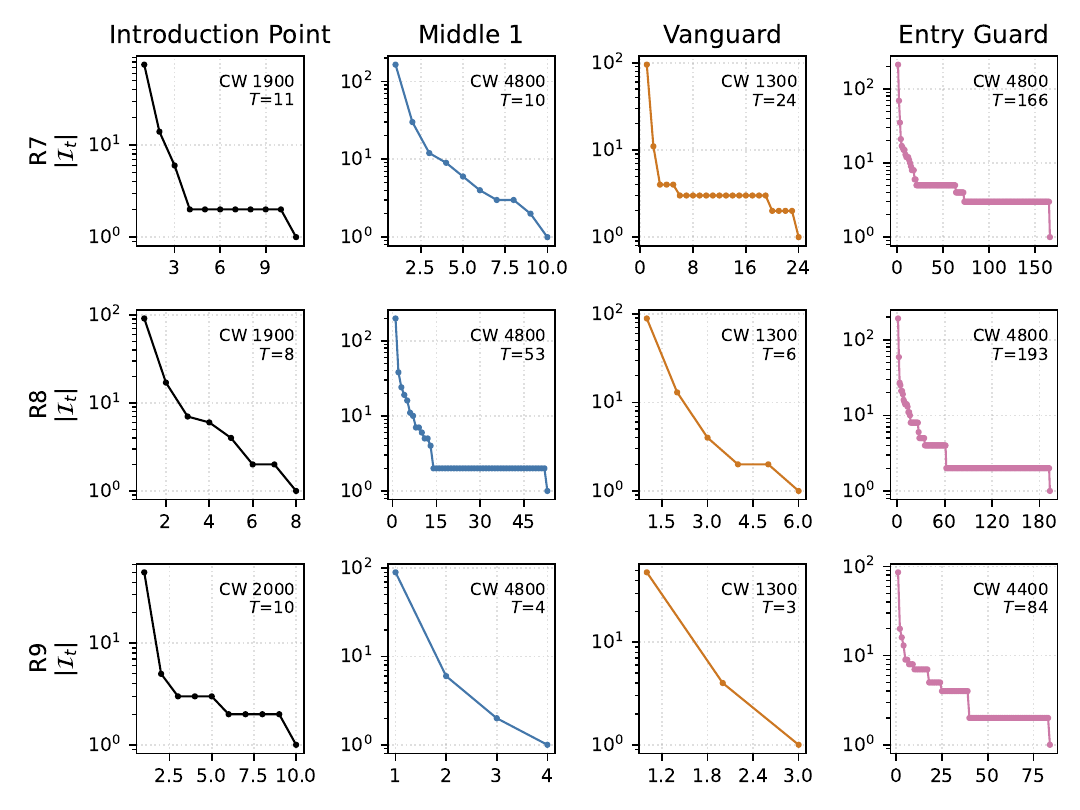}
\caption{Intersection degradation for Runs~7--9 across the four reconstruction
stages. Axes and annotations follow Figure~\ref{fig:runs-grid-a}.}
\label{fig:runs-grid-c}
\end{figure*}

This appendix reports the complete convergence record of
Section~\ref{sec:eval}. All $36$ individual run--stage observations are presented in
Table~\ref{tab:run-stage-thresholds} and the full intersection trajectory of
every run in Figures~\ref{fig:runs-grid-a}--\ref{fig:runs-grid-c}. 

\begin{table*}[tbp]
\centering
\scriptsize
\setlength{\tabcolsep}{5pt}
\caption{Per-run, per-stage measurements for nine end-to-end runs (IDs~1--9)
conducted on 7--10~January~2026.
$|\mathcal{A}_1|$ is the initial anonymity-set size;
$T_{\leq q}=\min\{t:|\mathcal{I}_t|\leq q\}$;
$T_{\mathrm{conv}}=\min\{t:|\mathcal{I}_t|=1\}$; and
\textbf{CW} is the stage-start consensus weight.}
\label{tab:run-stage-thresholds}
\begin{tabular}{@{}l l r rrrrr r@{}}
\toprule
\textbf{Run / UTC} & \textbf{Stage} & \textbf{$|\mathcal{A}_1|$} &
$T_{\leq 10}$ & $T_{\leq 5}$ & $T_{\leq 3}$ & $T_{\leq 2}$ &
$T_{\mathrm{conv}}$ & \textbf{CW} \\
\midrule
R1 (Day 1, 02:00)  & Intro. Point  & 79  & 3  & 4  & 42  & 245  & 285  & 830 \\
  & Middle 1  & 137  & 3  & 3  & 4  & 5  & 5  & 5000 \\
  & Vanguard  & 112  & 2  & 61  & 61  & 61  & 61  & 1200 \\
  & Entry Guard  & 218  & 8  & 16  & 16  & 31  & 31  & 9300 \\
\addlinespace

R2 (Day 1, 10:00)  & Intro. Point  & 101  & 2  & 3  & 9  & 52  & 52  & 850 \\
  & Middle 1  & 192  & 4  & 5  & 7  & 13  & 13  & 5000 \\
  & Vanguard  & 76  & 3  & 3  & 3  & 3  & 4  & 1200 \\
  & Entry Guard  & 284  & 10  & 17  & 17  & 21  & 62  & 9300 \\
\addlinespace

R3 (Day 1, 18:00)  & Intro. Point  & 74  & 3  & 9  & 11  & 163  & 163  & 850 \\
  & Middle 1  & 129  & 3  & 5  & 6  & 7  & 7  & 4000 \\
  & Vanguard  & 85  & 2  & 2  & 4  & 12  & 12  & 1200 \\
  & Entry Guard  & 249  & 11  & 25  & 26  & 26  & 78  & 9300 \\
\addlinespace

R4 (Day 2, 02:00)  & Intro. Point  & 81  & 3  & 4  & 4  & 4  & 8  & 850 \\
  & Middle 1  & 148  & 4  & 7  & 8  & 8  & 11  & 4000 \\
  & Vanguard  & 107  & 3  & 5  & 5  & 6  & 10  & 1200 \\
  & Entry Guard  & 256  & 13  & 21  & 45  & 49  & 123  & 9900 \\
\addlinespace

R5 (Day 2, 10:00)  & Intro. Point  & 101  & 6  & 11  & 30  & 178  & 178  & 1500 \\
  & Middle 1  & 158  & 6  & 10  & 13  & 16  & 17  & 4100 \\
  & Vanguard  & 148  & 38  & 60  & 61  & 76  & 142  & 1200 \\
  & Entry Guard  & 180  & 7  & 11  & 14  & 15  & 167  & 4800 \\
\addlinespace

R6 (Day 2, 18:00)  & Intro. Point  & 84  & 3  & 14  & 233  & 324  & 324  & 1400 \\
  & Middle 1  & 147  & 3  & 4  & 9  & 10  & 10  & 4100 \\
  & Vanguard  & 161  & 11  & 19  & 49  & 66  & 251  & 1200 \\
  & Entry Guard  & 98  & 7  & 14  & 23  & 177  & 185  & 4800 \\
\addlinespace

R7 (Day 3, 18:00)  & Intro. Point  & 75  & 3  & 4  & 4  & 4  & 11  & 1900 \\
  & Middle 1  & 164  & 4  & 6  & 7  & 9  & 10  & 4800 \\
  & Vanguard  & 95  & 3  & 3  & 6  & 20  & 24  & 1300 \\
  & Entry Guard  & 212  & 14  & 21  & 73  & 166  & 166  & 4800 \\
\addlinespace

R8 (Day 4, 02:00)  & Intro. Point  & 91  & 3  & 5  & 6  & 6  & 8  & 1900 \\
  & Middle 1  & 198  & 7  & 11  & 14  & 14  & 53  & 4800 \\
  & Vanguard  & 89  & 3  & 3  & 4  & 4  & 6  & 1300 \\
  & Entry Guard  & 191  & 16  & 28  & 62  & 62  & 193  & 4800 \\
\addlinespace

R9 (Day 4, 10:00)  & Intro. Point  & 50  & 2  & 2  & 3  & 6  & 10  & 2000 \\
  & Middle 1  & 90  & 2  & 3  & 3  & 3  & 4  & 4800 \\
  & Vanguard  & 48  & 2  & 2  & 3  & 3  & 3  & 1300 \\
  & Entry Guard  & 86  & 5  & 18  & 40  & 40  & 84  & 4400 \\
\bottomrule
\end{tabular}
\end{table*}

 \FloatBarrier
 \section{Prototype Implementation}
\label{app:implementation}

% Colocated variant of the coloured transit figure (Figure fig:protocol).
% Same onion-service protocol, but the introduction point IntP is repositioned
% next to the client and enclosed, together with the onion client, tcpdump and
% the intersection plugin, inside an "adversary host (same machine)" box.
% All protocol traffic still traverses Tor (colocation does not shortcut it).
%
% Layout mirrors fig-protocol-percircuit.tex: ONE entry guard per side (EG-S,
% EG-C), drawn larger than the layer-2 vanguards, with the endpoint-to-guard
% segments running as parallel lanes grouped by circuit (0.18 between the
% request and the acknowledgment of one circuit, 0.6+ between circuits), and the
% corner of every polyline hidden inside the guard disc.
% Requires: positioning, arrows.meta, calc, shapes.symbols, decorations.markings.
\definecolor{stepone}{HTML}{332288}   % 1  ESTABLISH_INTRO      indigo
\definecolor{steptwo}{HTML}{EE3377}   % 2  publish descriptor   magenta
\definecolor{stepthree}{HTML}{CC3311} % 6  fetch descriptor     red
\definecolor{stepfour}{HTML}{0077BB}  % 7  ESTABLISH_RENDEZVOUS blue
\definecolor{stepfive}{HTML}{33BBEE}  % 9  INTRODUCE1           cyan
\definecolor{stepsix}{HTML}{EE7733}   % 10 INTRODUCE2           orange
\definecolor{stepseven}{HTML}{117733} % 11 RENDEZVOUS1          green
\definecolor{stepeight}{HTML}{009988} % 12 RENDEZVOUS2          teal
\definecolor{stepnine}{HTML}{444444}  % 14 application data     grey
\definecolor{obscol}{HTML}{CC3311}    % adversary instrumentation (red)
\definecolor{startcol}{HTML}{009E73}  % 8 / 13 START and STOP signals (green)
% Intersection plugin icon (self-contained; \providecommand so it does not clash
% with the copy in fig-intersection-system.tex).
\providecommand{\intersectionpluginicon}[1][12mm]{%
  \resizebox{!}{#1}{%
  \begin{tikzpicture}[x=1mm,y=1mm,line cap=round,line join=round,
      ipk/.style={draw=black,fill=white,line width=0.45mm,rounded corners=0.8mm},
      ips/.style={draw=black,line width=0.45mm}]
    \draw[ipk]
      (1,1) -- (9,1)
      -- (9,3.4) arc (270:90:1.6)
      -- (9,9)
      -- (6.6,9) arc (0:180:1.6)
      -- (1,9) -- cycle;
    \draw[ips] (3.5,2.8) -- (3.5,5.1) arc (180:0:1.45) -- (6.4,2.8);
\end{tikzpicture}%
  }%
}
\begin{figure*}[!t]
\centering
\resizebox{\linewidth}{!}{%
\begin{tikzpicture}[
  >={Latex[length=3.2mm,width=2.5mm]},
  cas/.style={preaction={draw,line width=4.4pt,white,rounded corners=3pt}},
  req/.style={->,line width=2.3pt,rounded corners=3pt,cas},
  fwd/.style={->,line width=2.3pt,dash pattern=on 3.6pt off 2.4pt,rounded corners=3pt,cas},
  ackl/.style={->,line width=1.8pt,densely dotted,preaction={draw,line width=3.5pt,white}},
  appl/.style={{Latex[length=3.2mm,width=2.6mm]Latex[length=3.2mm,width=2.6mm,reversed,sep=2.4mm]}-{Latex[length=3.2mm,width=2.6mm,reversed,sep=2.4mm]Latex[length=3.2mm,width=2.6mm]},line width=2.5pt,rounded corners=3pt,cas},
  rb/.style n args={2}{circle,fill=#1,text=#2,font=\normalsize\bfseries,inner sep=0.3pt,minimum size=6.2mm,preaction={draw=white,line width=3pt}},
  abd/.style={circle,draw=#1,dash pattern=on 0.4pt off 1.4pt,line width=0.9pt,fill=white,text=#1!80!black,font=\normalsize\bfseries,inner sep=0.3pt,minimum size=6.2mm,preaction={draw=white,line width=2.5pt}},
  stn/.style={fill=white,draw=gray!45,line width=0.5pt},
  gstn/.style={fill=white,draw=gray!55,line width=0.9pt},
  fstn/.style={fill=white,draw=gray!35,dash pattern=on 1.2pt off 1.2pt,line width=0.5pt},
  rlab/.style={font=\large\bfseries,inner sep=1.5pt,fill=white,rounded corners=1pt},
  elab/.style={font=\large\bfseries,align=center,inner sep=1.5pt},
  hostbox/.style={draw=black!78,line width=1.4pt,rounded corners=10pt,fill=black!3},
  hlab/.style={font=\footnotesize\bfseries,align=center,inner sep=1.2pt}
]
% ================ Tor network cloud ================
\node[cloud, cloud puffs=21, cloud puff arc=104, aspect=1.70, draw=gray!30,
      line width=0.4pt, fill=gray!7, minimum width=33.00cm, minimum height=19.60cm]
      at (-2.30,0.45) {};
% ================ adversary host box (behind arrows and icons) ================
% Encloses EXACTLY {Client, IntP, tcpdump, intersection plugin, controller}.
% EG-C and every VG-C / M-C relay stays OUTSIDE.
\draw[hostbox] (-3.85,-3.85) rectangle (3.85,3.85);
\node[hlab,anchor=south west,fill=black!3,inner sep=1.5pt] at (-3.68,-3.72)
      {adversary host (same machine)};
% ================ protocol strokes (parallel lanes, generated) ================
% ---- node coordinates ----
\coordinate (hs) at (-18.00,0.00);
\coordinate (egS) at (-14.00,0.00);
\coordinate (vgS1) at (-10.20,4.40);
\coordinate (vgS2) at (-10.20,1.40);
\coordinate (vgS3) at (-10.20,-1.70);
\coordinate (vgS4) at (-10.20,-4.90);
\coordinate (mSd) at (-6.80,6.85);
\coordinate (mSi) at (-6.60,1.70);
\coordinate (mSr) at (-6.00,-6.80);
\coordinate (hd) at (0.00,9.40);
\coordinate (rp) at (0.00,-7.80);
\coordinate (ip) at (-1.55,1.70);
\coordinate (cl) at (1.55,1.70);
\coordinate (egC) at (7.60,1.70);
\coordinate (vgC1) at (11.90,4.60);
\coordinate (vgC2) at (12.40,1.30);
\coordinate (vgC3) at (4.60,7.00);
\coordinate (vgC4) at (12.40,-2.10);
\coordinate (vgR) at (10.60,-5.00);
\coordinate (mCd) at (8.00,8.20);
\coordinate (mC2) at (-1.55,6.60);
\coordinate (mCr) at (5.00,-6.80);
% ---- protocol strokes (parallel lanes, grouped by circuit) ----
\draw[req,steptwo] (-16.850,1.000) -- (-14.500,1.000) -- (-13.931,0.601) -- (-10.431,4.652) -- (-6.962,7.152) -- (-0.887,9.430);
\draw[ackl,steptwo] (-0.648,8.794) -- (-6.638,6.548) -- (-9.969,4.148) -- (-13.416,0.156) -- (-14.500,0.820) -- (-16.850,0.820);
\draw[req,stepone] (-16.850,0.200) -- (-14.500,0.200) -- (-13.669,0.548) -- (-10.287,1.794) -- (-6.617,2.100) -- (-2.750,2.100);
\draw[ackl,stepone] (-2.750,1.700) -- (-6.600,1.700) -- (-10.200,1.400) -- (-13.531,0.173) -- (-14.500,0.000) -- (-16.850,0.000);
\draw[fwd,stepsix] (-2.750,1.300) -- (-6.583,1.300) -- (-10.113,1.006) -- (-13.393,-0.202) -- (-14.500,-0.200) -- (-16.850,-0.200);
\draw[req,stepseven] (-16.850,-1.040) -- (-14.500,-1.040) -- (-13.962,-0.603) -- (-10.417,-5.175) -- (-6.100,-7.128) -- (-0.950,-8.580);
\draw[req,stepthree] (2.400,1.900) -- (7.100,1.900) -- (7.824,2.261) -- (11.354,4.642) -- (7.846,7.879) -- (0.760,8.942);
\draw[ackl,stepthree] (0.861,9.615) -- (8.154,8.521) -- (12.446,4.558) -- (8.205,1.698) -- (7.100,1.700) -- (2.400,1.700);
\draw[req,stepfive] (2.400,2.520) -- (7.549,2.520) -- (4.802,7.374) -- (-1.910,6.937) -- (-1.910,2.900);
\draw[ackl,stepfive] (-1.190,2.900) -- (-1.190,6.263) -- (4.398,6.626) -- (6.620,2.700) -- (2.400,2.700);
\draw[req,stepfour] (2.400,1.040) -- (7.100,1.040) -- (8.270,1.452) -- (11.298,-5.311) -- (5.128,-7.294) -- (1.257,-8.069);
\draw[ackl,stepfour] (1.190,-7.735) -- (5.043,-6.965) -- (10.833,-5.104) -- (7.959,1.313) -- (7.100,0.860) -- (2.400,0.860);
\draw[req,stepeight] (1.124,-7.402) -- (4.957,-6.635) -- (10.367,-4.896) -- (7.649,1.174) -- (7.100,0.680) -- (2.400,0.680);
\draw[appl,stepnine] (-16.850,-0.860) -- (-14.500,-0.860) -- (-13.425,-0.187) -- (-9.983,-4.625) -- (-5.900,-6.472) -- (0.000,-7.250) -- (4.872,-6.306) -- (9.902,-4.689) -- (7.339,1.035) -- (7.100,0.500) -- (2.400,0.500);
% re-stroke the host-box border: the strokes' white casing erases it where the
% client lanes cross, so draw the outline again (no fill) on top of them.
\draw[hostbox,fill=none] (-3.85,-3.85) rectangle (3.85,3.85);

% ---- numbered badges ----
\node[rb={steptwo}{white}] at (-12.163,2.647) {2a};
\node[rb={stepone}{white}] at (-12.124,1.117) {1a};
\node[rb={stepsix}{white}] at (-8.260,1.160) {10};
\node[rb={stepseven}{white}] at (-12.494,-2.496) {11};
\node[rb={stepthree}{white}] at (9.375,3.307) {6a};
\node[rb={stepfive}{black}] at (5.992,5.271) {9a};
\node[rb={stepfour}{white}] at (9.729,-1.808) {7a};
\node[rb={stepeight}{white}] at (7.428,-5.841) {12};
\node[rb={stepnine}{white}] at (-3.301,-6.815) {14};
\node[abd=steptwo] at (-8.530,5.185) {2b};
\node[abd=stepone] at (-3.799,1.700) {1b};
\node[abd=stepthree] at (9.829,6.974) {6b};
\node[abd=stepfive] at (6.208,3.428) {9b};
\node[abd=stepfour] at (9.827,-2.858) {7b};
% ================ relays ======================================================
% ONE entry guard per side: a client or a service keeps a persistent set of three
% guards but sends every circuit through the first reachable one, so a single
% guard carries all of that side's traffic. Drawn larger (14mm vs 10mm) to mark
% the shared hop, exactly as in Figure~\ref{fig:protocol-percircuit}.
\foreach \c/\l/\dx in {egS/EG-S/0,egC/EG-C/-1.05}{%
  \draw[gstn] (\c) circle (14mm);
  \node[inner sep=0] at (\c) {\includegraphics[height=13mm]{tor_figure/router.png}};
  \node[rlab] at ($(\c)+(\dx,-1.78)$) {\l};}
% layer-2 vanguards: a persistent set of four per side, one per circuit; the
% members no circuit in this figure uses are drawn faded.
\foreach \c/\l in {vgS1/VG-S-1,vgS2/VG-S-2,vgS4/VG-S-4,vgC1/VG-C-1,vgR/VG-C-2}{%
  \draw[stn] (\c) circle (10mm);
  \node[inner sep=0] at (\c) {\includegraphics[height=9.5mm]{tor_figure/router.png}};
  \node[rlab] at ($(\c)+(0,-1.28)$) {\l};}
% VG-C-3 sits above the host box; its label goes down-left, clear of both strokes
\draw[stn] (vgC3) circle (10mm);
\node[inner sep=0] at (vgC3) {\includegraphics[height=9.5mm]{tor_figure/router.png}};
\node[rlab] at ($(vgC3)+(-1.55,-0.95)$) {VG-C-3};
\foreach \c/\l in {vgS3/VG-S-3,vgC4/VG-C-4}{%
  \draw[fstn] (\c) circle (10mm);
  \node[inner sep=0,opacity=0.45] at (\c) {\includegraphics[height=9.5mm]{tor_figure/router.png}};
  \node[rlab,text=gray!60] at ($(\c)+(0,-1.28)$) {\l};}
% middle relays, one per circuit
\foreach \c/\l in {mSd/M-S-d,mSi/M-S-i,mSr/M-S-r,mCd/M-C-1,mCr/M-C-r}{%
  \draw[stn] (\c) circle (11.4mm);
  \node[inner sep=0] at (\c) {\includegraphics[height=11.6mm]{tor_figure/router.png}};
  \node[rlab] at ($(\c)+(0,-1.34)$) {\l};}
% the client's introduction middle sits above the host box; its label goes on top
\draw[stn] (mC2) circle (11.4mm);
\node[inner sep=0] at (mC2) {\includegraphics[height=11.6mm]{tor_figure/router.png}};
\node[rlab] at ($(mC2)+(-1.95,0.05)$) {M-C-2};
% monitored introduction point (inside the host box) and rendezvous point
\draw[stn] (ip) circle (10.4mm);
\node[inner sep=0] at (ip) {\includegraphics[height=11.6mm]{tor_figure/router.png}};
\node[rlab] at ($(ip)+(0,-1.34)$) {IntP};
\draw[stn] (rp) circle (10.4mm);
\node[inner sep=0] at (rp) {\includegraphics[height=11.6mm]{tor_figure/router.png}};
\node[rlab] at ($(rp)+(0,1.45)$) {RP};
% ================ endpoints and HSDir ========================================
\fill[white] (hs) circle (11.0mm);
\node[inner sep=0] at (hs) {\includegraphics[height=22mm]{tor_figure/server.png}};
\node[elab] at ($(hs)+(0,-1.85)$) {Onion\\Service};
\fill[white] (cl) circle (7mm);
\node[inner sep=0] at (cl) {\includegraphics[height=22mm]{tor_figure/client.png}};
\node[elab] at ($(cl)+(0,1.42)$) {Client};
\fill[white] (hd) circle (7.0mm);
\node[inner sep=0] at (hd) {\includegraphics[height=12mm]{tor_figure/hs_authority.png}};
\node[elab] at ($(hd)+(0,1.10)$) {HSDir};
% ================= colocated instruments (inside the host box) ================
\node[inner sep=0] (tcpn) at (-2.45,-2.20) {\includegraphics[height=11mm]{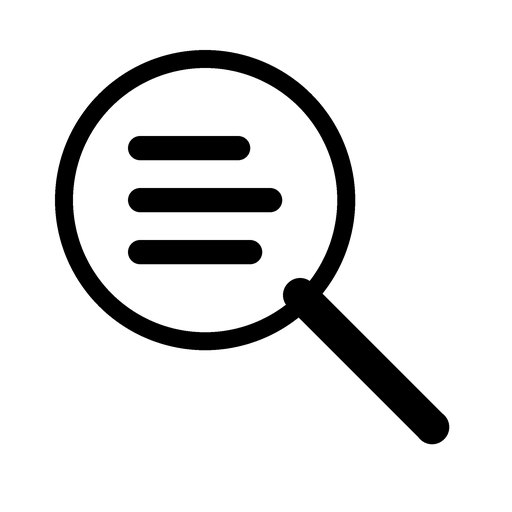}};
\node[hlab,below=0.6mm of tcpn] {tcpdump};
\node[inner sep=0] (plugn) at (1.60,-2.20) {\intersectionpluginicon[11mm]};
\node[hlab,below=0.6mm of plugn] {intersection plugin};
% controller / trial automation (runner.sh)
\begin{scope}[shift={(-2.70,-0.55)}]
  \draw[line width=1pt,rounded corners=1.4pt,fill=black!82] (-0.52,-0.34) rectangle (0.52,0.34);
  \draw[line width=1.1pt,white,line cap=round,line join=round] (-0.34,0.13) -- (-0.20,0.00) -- (-0.34,-0.13);
  \draw[line width=1.1pt,white,line cap=round] (-0.10,-0.15) -- (0.20,-0.15);
\end{scope}
\node[hlab] (ctrln) at (-2.70,-1.15) {controller};
% ---- 3: the controller launches the onion client and the intersection plugin
\draw[-{Latex[length=3.0mm,width=2.4mm]},dashed,line width=1.2pt,black!70]
      (-2.12,-0.32) -- (0.76,0.62);
\draw[-{Latex[length=3.0mm,width=2.4mm]},dashed,line width=1.2pt,black!70]
      (-2.12,-0.80) -- node[rb={black!70}{white},pos=.52]{3} (1.02,-1.80);
% ---- 5: tcpdump streams the captured headers continuously to the plugin
\draw[-{Latex[length=3.2mm,width=2.6mm]},line width=1.6pt,black!62,
      postaction={decorate,decoration={markings,
        mark=between positions 0.12 and 0.84 step 0.18 with {\arrow[black!62]{Latex[length=2.2mm,width=1.9mm]}}}}]
      (-1.85,-2.14) -- node[rb={black!70}{white},pos=.30]{5} (1.05,-2.14);
% ---- 4: the intersection plugin launches tcpdump as a subprocess
\draw[-{Latex[length=3.0mm,width=2.4mm]},dashed,line width=1.2pt,black!70]
      (1.05,-2.64) -- node[rb={black!70}{white},pos=.42]{4} (-1.85,-2.64);
% ---- 8 START / 13 STOP: the client opens and closes the capture window
\draw[-{Latex[length=4.2mm,width=3.4mm]},line width=2.3pt,startcol]
      (1.22,0.36) -- node[rb={startcol}{white},pos=.38]{8} (1.22,-1.70);
\draw[-{Latex[length=4.2mm,width=3.4mm]},line width=2.3pt,startcol]
      (2.28,0.24) -- node[rb={startcol}{white},pos=.50]{13} (2.28,-1.32) -- (1.98,-1.70);
\end{tikzpicture}%
}

\caption{The onion-service protocol with the colocated adversary overlaid. The
probe client, \texttt{tcpdump}, the intersection plugin, and the monitored
Introduction Point (IntP) share one host (the box). Colocation does not shortcut
Tor: \texttt{ESTABLISH\_INTRO}~(1a), \texttt{INTRODUCE1}~(9a), and
\texttt{INTRODUCE2}~(10) still traverse the pinned four-hop circuits. Steps
1a--2b set up the service, 3--5 bring up the capture at each stage, and 6a--14
form one iteration. Node and badge conventions follow
Figure~\ref{fig:protocol-percircuit}.}
\label{fig:protocol-colocated}

\end{figure*}

This appendix describes the prototype used in
Section~\ref{sec:experimental-setup}, for which we modify the Tor daemons of our
onion service and probe client. Figure~\ref{fig:protocol-colocated} summarizes
it and numbers the steps referenced below.

\paragraph{Components.}
Four of the five components (Table~\ref{tab:impl-components}) run on the
monitored relay's host: a controller, a modified probe-client daemon, an
intersection plugin, and a \texttt{tcpdump} subprocess; the fifth is the
modified onion-service daemon on the service host. The four move with the
observation at each stage advance (below).

\begin{table}[tb]
\footnotesize
\setlength{\tabcolsep}{4pt}
\caption{Prototype components and their roles.}
\label{tab:impl-components}
\begin{tabular}{@{}>{\raggedright\arraybackslash}p{0.25\linewidth}
                    >{\raggedright\arraybackslash}p{0.67\linewidth}@{}}
\toprule
\textbf{Component} & \textbf{Role} \\
\midrule
Controller &
Runs iterations and advances observation to the identified successor. \\

Probe-client daemon &
Triggers one introduction handshake per iteration and signals the beginning
and end of the observation window. \\

Intersection plugin &
Constructs anonymity sets from observed destination addresses and maintains
their cumulative intersection. \\

\texttt{tcpdump} &
Captures the headers of the monitored relay's outgoing IPv4 packets. \\

Onion-service daemon &
Pins the target introduction circuit to the four relays we operate. \\
\bottomrule
\end{tabular}
\end{table}

\paragraph{Service setup: pinning the introduction circuit (1a--2b).}
Every onion service establishes introduction points and publishes a descriptor
before it is reachable (Section~\ref{sec:onion-service}). We leave that procedure
intact and change only the relays it runs over: for experimental validation, and
to obtain visibility at each monitored relay, we instrument our service to
construct its introduction circuit~(\textbf{1a},~\textbf{1b}) in the order
\[
  \underbrace{%
    \textnormal{entry guard}\rightarrow
    \textnormal{layer-2 vanguard}\rightarrow
    \textnormal{middle relay}\rightarrow
    \textnormal{Introduction Point}%
  }_{\textnormal{four public relays we operate}}.
\]
The service publishes its signed descriptor to the responsible
HSDirs~(\textbf{2a},~\textbf{2b}), making it reachable from the public
directory. Operating the four relays ourselves substitutes for the visibility an
adversary would otherwise have to obtain by other means by a state adversary. The four names label positions on
our circuit rather than properties of the relays: our Introduction Point may at
the same time be another user's entry guard. The four monitored-relay stages walk this circuit in reverse, after
which the final observation identifies the onion service's network endpoint.

\paragraph{Once per stage: bring-up (3--5).}
At stage~$0$ the monitored relay is the advertised Introduction Point. At the
beginning of each stage, the controller starts the intersection
plugin~(\textbf{3}), which spawns its \texttt{tcpdump} subprocess on the
monitored relay~(\textbf{4}). Packet headers reach the plugin through a
pipe~(\textbf{5}) rather than a capture file. The plugin generates a fresh RSA
pseudonymization key for the stage; each stage therefore runs a fresh plugin
process, and the key dies with it.

\paragraph{Once per iteration: one handshake and the intersection fold (6a--14).}
 Each iteration~$j$ begins with the controller
launching a fresh probe client against our onion service. Having no cached
state, the client fetches the descriptor from the responsible
HSDirs~(\textbf{6a},~\textbf{6b}), learns the Introduction Point of the pinned
circuit, and registers with a rendezvous point~(\textbf{7a},~\textbf{7b}). The
modified client signals \texttt{START}~(\textbf{8}) to the plugin immediately
before sending \texttt{INTRODUCE1}~(\textbf{9a},~\textbf{9b}). From then until
\texttt{STOP}, the plugin reads the destination IP address of each header,
pseudonymizes it under the stage key, and marks it as visited in an in-memory
map. The Introduction Point forwards the request along the pinned circuit as
\texttt{INTRODUCE2}~(\textbf{10}); the service builds its own circuit to the
rendezvous point and sends \texttt{RENDEZVOUS1}~(\textbf{11}); the rendezvous
point joins the two circuits and returns \texttt{RENDEZVOUS2} to the
client~(\textbf{12}), on receipt of which the client signals
\texttt{STOP}~(\textbf{13}), closing the window. The service then serves the
requested page over the joined rendezvous circuits~(\textbf{14}), after
\texttt{STOP} and therefore unobserved.

The pseudonyms marked between \texttt{START} and \texttt{STOP} form
$\mathcal{A}_i^{(j)}$. At \texttt{STOP} the plugin folds this set into the
cumulative intersection according to Algorithm~\ref{alg:reconstruction}, first
removing the known predecessor $r_{m_{i-1}}$ from stage~$1$ onward
(Section~\ref{sec:progressive-reconstruction}). If the intersection still holds
more than one pseudonym, the controller kills the probe client, sleeps
$\delta=30$~s, and repeats steps~6a--13 against the same relay; the client
launched for the next iteration therefore starts with no cached state.

\paragraph{Stage advance and termination.}
If the cumulative intersection holds a single pseudonym, the stage is done: the
controller moves the observation components over SSH to that relay and repeats
steps~3--13 there, so that the plugin restarts with a fresh stage key and an
empty intersection. If the intersection is empty, the pinned path no longer
holds and the run ends with the error \textsc{IntroductionCircuitDropped}.

\paragraph{Ground truth.}
Knowledge of the pinned path is used only for validation: it is the ground truth
for the successor recovered at each stage, under the criteria given in
Section~\ref{sec:experimental-setup}. The validation check reads
$\mathcal{A}_i^{(j)}$ and $\mathcal{I}_i^{(j)}$ after the fold and writes to
neither. Ground-truth information is never used to construct anonymity sets,
remove candidates, or determine convergence.

\paragraph{Fidelity to the threat model.}
Section~\ref{sec:anonymity-set} gives the adversary timestamped headers of the
monitored relay's outgoing packets together with the times at which its own
client sent \texttt{INTRODUCE1} and received \texttt{RENDEZVOUS2}, and has it
keep the headers falling between the two. Both handshake events occur at the
adversary's own client, so raising them in band as \texttt{START} and
\texttt{STOP} (steps~8 and~13) rather than recording their times tells it
nothing new: signals (START,STOP) now bound $W_i^{(j)}$, and the clock agreement that
Section~\ref{sec:anonymity-set} assumes holds by construction. Because
\texttt{START} is raised immediately before the cell leaves the client, the
delimited interval is a slight superset of $W_i^{(j)}$, which can only admit
further candidates into $\mathcal{A}_i^{(j)}$ and never drop the successor.
Visibility at the monitored relay is the one capability colocation does not
supply; it comes from operating the relays ourselves, as noted above.
Colocation is a convenience of our deployment: what it buys is that clock
alignment, not any relaxation of what the adversary must be able to see; a
timestamp on a captured header would have no remaining use.

Only set membership is retained: packet ordering, multiplicity, and
packet-level timestamps are discarded. Raw addresses, pseudonyms, anonymity
sets, intersections, and pseudonymization keys remain only in volatile memory
and are destroyed when the plugin exits at each stage advance. The prototype
writes only aggregate experimental data: one CSV records the run, stage,
iteration, and intersection cardinality per observation, a second records
stage-level metadata, including the monitored relay's public Onionoo metrics and
experiment timestamp, and nothing else reaches disk.

\subsection{Ethics Considerations}
\label{sec:ethics}

This study was designed and conducted in accordance with the guidelines of the Tor Research Safety Board (TRSB)\footnote{\href{https://research.torproject.org/safetyboard.html}{Tor Research Safety Board}}. We considered each of the TRSB's nine research-safety principles as follows.
\begin{enumerate}
    \item \textbf{Use a test Tor network whenever possible.}
    We considered conducting the experiments in a private testbed or simulated Tor network. However, our attack depends on realistic background traffic and relay activity. A typical ten-minute period on the deployed network involves roughly 550{,}000 active users and about 1.4~million active circuits~\cite{jansen2016safely}, while existing testbeds and simulation or emulation frameworks operate at substantially smaller scale and cannot jointly reproduce the deployed network's relay population, routing behaviour, bandwidth distribution, congestion, and workloads~\cite{shirazi2015tor}. Because downscaling these properties would directly affect the intersection behaviour being evaluated, we used the live Tor network to demonstrate the feasibility of our attack under realistic conditions.

    \item \textbf{Attack only yourself or your own traffic.}
    All deanonymization experiments targeted only the introduction circuit of our self-operated onion service. We constructed and intersected anonymity sets only for this circuit and never attempted to deanonymize third-party circuits, services, or clients. Third-party Tor traffic appeared only as naturally occurring background traffic at our relays.

    \item \textbf{Collect only data that are safe to make public.}
    Persisted experimental results contained only aggregate intersection sizes, trial indices, stage-start UTC timestamps, and public Onionoo relay metadata. No raw or pseudonymised destination addresses, packet traces, or payloads were retained. The resulting dataset therefore contains no information that, on its own, identifies unrelated Tor users or recovers the identity of a third party whose pseudonym temporarily appeared in an anonymity set.

    \item \textbf{Do not collect data that are not needed.}
    We only collected data during each client-defined \texttt{INTRODUCE1}--\texttt{RENDEZVOUS2} window, and we processed only the destination IPv4 addresses required to construct the anonymity set. We did not retain packet payloads, stream identifiers, packet ordering, traffic multiplicity, or per-packet timestamps and any data outside the observation windows.

    \item \textbf{Take reasonable security precautions.}
    Destination IPv4 addresses required during processing were pseudonymised in volatile memory using per-stage RSA keys that never left process memory. Communication between \texttt{tcpdump} and the intersection plugin used pipes, which rely on volatile kernel-memory buffers and do not write their contents to disk. Intermediate observations, anonymity sets and cryptographic material remained in volatile memory and were destroyed when the corresponding processes exited.

    \item \textbf{Limit the granularity of collected data.}
    Our colocation methodology allowed the intersection plugin to process only observations falling within the \texttt{INTRODUCE1}--\texttt{RENDEZVOUS2} interval without collecting their packet-level timestamps. Observations were reduced immediately to set membership of pseudonymised destination addresses and ultimately to aggregate intersection cardinalities and relay metadata.

    \item \textbf{Ensure that the benefits outweigh the risks.}
    The residual risk of processing third-party background traffic was limited by the safeguards described above. In return, the live-network experiment demonstrates that long-lived introduction circuits can expose repeatable traffic patterns that allow an adversary to determine where to observe next, contrary to Tor's goal of limiting traffic analysis under partial observation. We additionally propose a mitigation that, assuming Tor's short-lived rendezvous circuits are resistant to this attack class, periodically rebuilds the internal introduction path to provide comparable protection.

    \item \textbf{Consider auxiliary data when assessing risk.}
    We considered whether retained measurements could become identifying when combined with external information. We therefore excluded destination addresses, pseudonyms, packet-level timestamps, ordering information, and traffic traces from the persistent dataset. The retained Onionoo data consist solely of public relay metadata and cannot restore these discarded observations.

    \item \textbf{Consider whether users intended their data to be private.}
    We treated all unrelated Tor traffic as privacy-sensitive. Such traffic was used solely as naturally occurring background activity and was never targeted for deanonymization. We made no attempt to associate observed destinations with users, services, or circuits.
\end{enumerate}
 \FloatBarrier
 \section{Introduction-Handshake Latency Measurements}
\label{app:introduction-latency}

In the paper, we argue that the introduction handshake executes within a short interval because only a small amount of data is transferred during the exchange. Consequently, the short observation window bounds the number of unrelated destination IP addresses that can be observed at the monitored relay and enter the anonymity set. Rather than separately modeling factors such as network RTTs, bottlenecks, relay bandwidth, and congestion, we evaluate the resulting handshake duration empirically under real Tor network conditions. Specifically, we measured the introduction handshake for 16 publicly reachable onion services from a residential computer, repeating the measurement 10 times for each service. We report the average \texttt{INTRODUCE1}--\texttt{RENDEZVOUS2} duration for each service below.
\paragraph{Methodology.}
We measure this interval using 16 publicly reachable onion services. For each service, we perform 10 independent introduction handshakes, yielding 160 measurements in total. For every trial, the client records the time immediately before sending \texttt{INTRODUCE1} and the time at which the corresponding \texttt{RENDEZVOUS2} is received. We define the measured duration as
\[
    \Delta t =
    t_{\texttt{RENDEZVOUS2}}
    -
    t_{\texttt{INTRODUCE1}}.
\]
Thus, $\Delta t$ captures the client-observed time required for the introduction request to traverse the client-side introduction circuit and the service-side introduction circuit, for the service to establish its rendezvous circuit, and for the resulting rendezvous handshake to return to the client. These measurements are used to empirically support our claim that the introduction handshake executes within a short time interval; the measured onion services are not targets of our deanonymization experiments.

\paragraph{Results.}
Table~\ref{tab:onion_avg_latency} reports the mean $\Delta t$ across the 10 trials for each onion service. The per-service means range from $0.521$~s to $1.780$~s. Despite involving both the introduction and rendezvous portions of the protocol, the complete interval therefore remains on the order of seconds.

\begin{table}[htbp]
\centering
\renewcommand{\arraystretch}{1.1}
\setlength{\tabcolsep}{5pt}
\footnotesize
\caption{Mean \texttt{INTRODUCE1}--\texttt{RENDEZVOUS2} interval across 16 onion services, with 10 trials per service.}
\label{tab:onion_avg_latency}
\begin{tabular}{l l c}
\hline
\textbf{Name} & \textbf{Type} & \textbf{Avg.\ $\Delta t$ (s)} \\
\hline
Cryptostamps     & Postage store        & 0.642 \\
Breaking Bad     & Drug forum           & 0.547 \\
Black Cloud      & Onion pastebin       & 1.780 \\
Ahmia            & HS search engine     & 1.019 \\
Onion ID Serv.   & ID/passport store    & 0.879 \\
ChaTor           & Onion messenger      & 1.230 \\
Comic Book Libr. & Library              & 0.654 \\
Apples4Bitcoin   & Onion apple store    & 0.882 \\
Dread            & Onion forum          & 1.060 \\
Mail2Tor         & Onion mail           & 0.701 \\
Sonar            & Onion messenger      & 0.805 \\
FAH              & Hiring service       & 1.492 \\
Mobile Store     & Mobile store         & 0.871 \\
USJUD            & Counterfeit store    & 0.636 \\
BMG              & Gun store            & 0.521 \\
DarkSearch       & Onion search engine  & 0.657 \\
\hline
\end{tabular}
\end{table}

%\paragraph{Relevance to the attack.}
%These measurements show that the attack's observation window is short relative to the lifetime of the target introduction circuit. During each window, the monitored relay has only a limited interval in which unrelated outgoing destinations can enter $\mathcal{A}_i^{(j)}$. Repeating these short observations against the same long-lived introduction circuit preserves the true successor across anonymity sets while allowing transient background destinations to vary and be removed by intersection. The measurements therefore provide empirical support for using the \texttt{INTRODUCE1}--\texttt{RENDEZVOUS2} interval as a narrow observation window in our evaluation.
 \FloatBarrier
\section{Jurisdictional Concentration and Obtaining Visibility}
\label{app:jurisdiction-concentration}

This appendix quantifies the geographic concentration of Tor relays and their path-selection probability mass. We consider the Fourteen Eyes as a grouping of jurisdictions with documented signals-intelligence cooperation~\cite{williams2023five} and aggregate Tor relay infrastructure accordingly. The security implications of this concentration for the feasibility of sequential network observation are discussed below.
\paragraph{Data collection.}
We obtain a snapshot of the public Tor relay network from Onionoo on 30 November 2025. For each relay, we record its reported hosting country and consensus-derived \texttt{guard\_probability} and \texttt{middle\_probability}. We classify a relay as being within the Fourteen Eyes when its reported hosting country is Australia, Belgium, Canada, Denmark, France, Germany, Italy, the Netherlands, New Zealand, Norway, Spain, Sweden, the United Kingdom, or the United States. All remaining hosting countries are grouped as outside the Fourteen Eyes.

\begin{figure}[!t]
    \centering
    \includegraphics[width=0.85\textwidth]{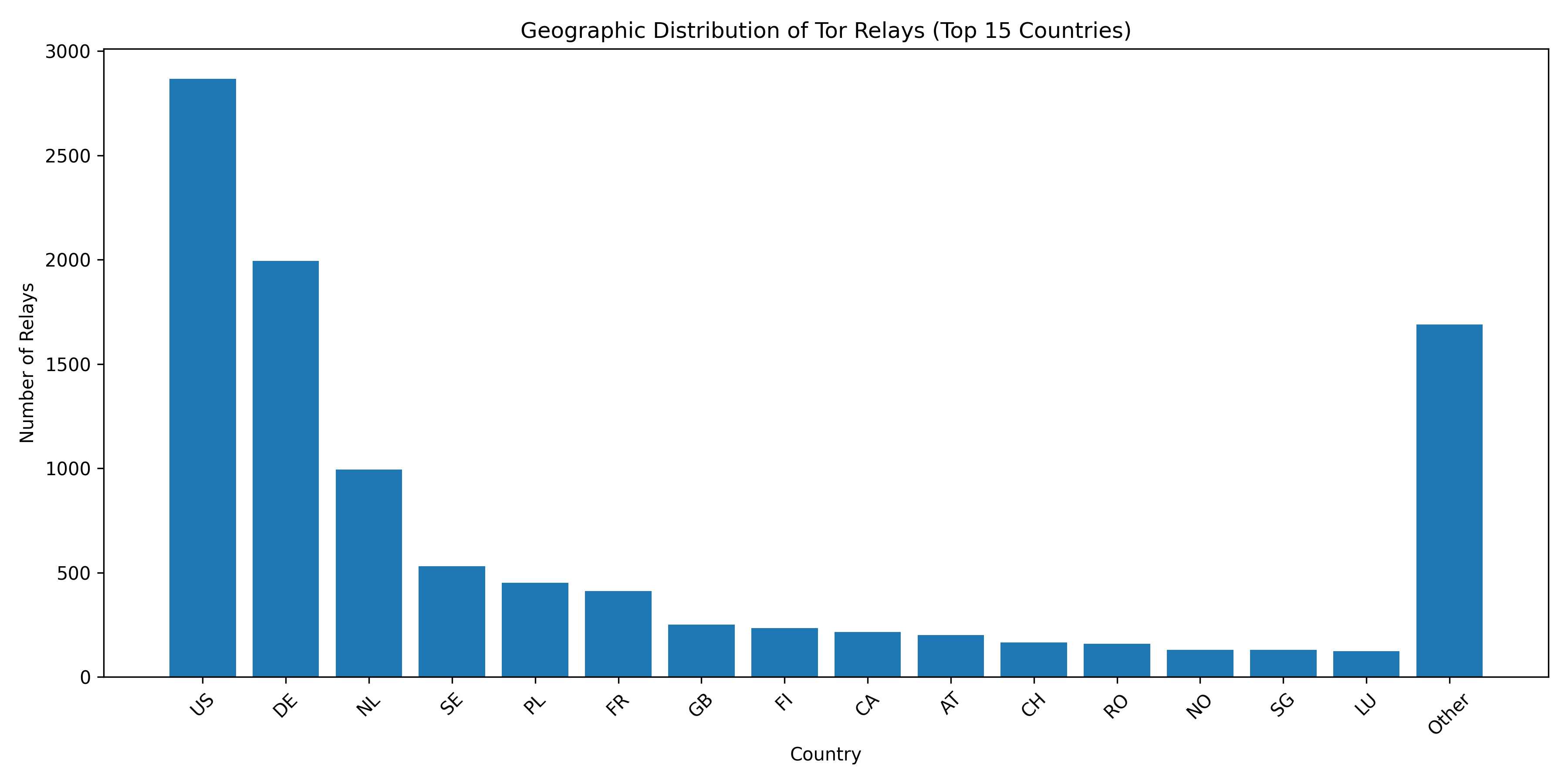}
    \caption{Geographic distribution of Tor relays in the 30 November 2025 Onionoo snapshot. Countries are distinguished according to whether they belong to the Fourteen Eyes.}
    \label{fig:country_distribution}
\end{figure}

\paragraph{Relay-count concentration.}
We first aggregate relays by their reported hosting country. Figure~\ref{fig:country_distribution} shows that Tor relay hosting is geographically concentrated rather than uniformly distributed. Germany, the United States, and the Netherlands account for a substantial share of the observed relay population. In aggregate, Fourteen-Eyes jurisdictions host more relays than all remaining jurisdictions combined.

\begin{figure}[!t]
    \centering
    \includegraphics[width=0.60\textwidth]{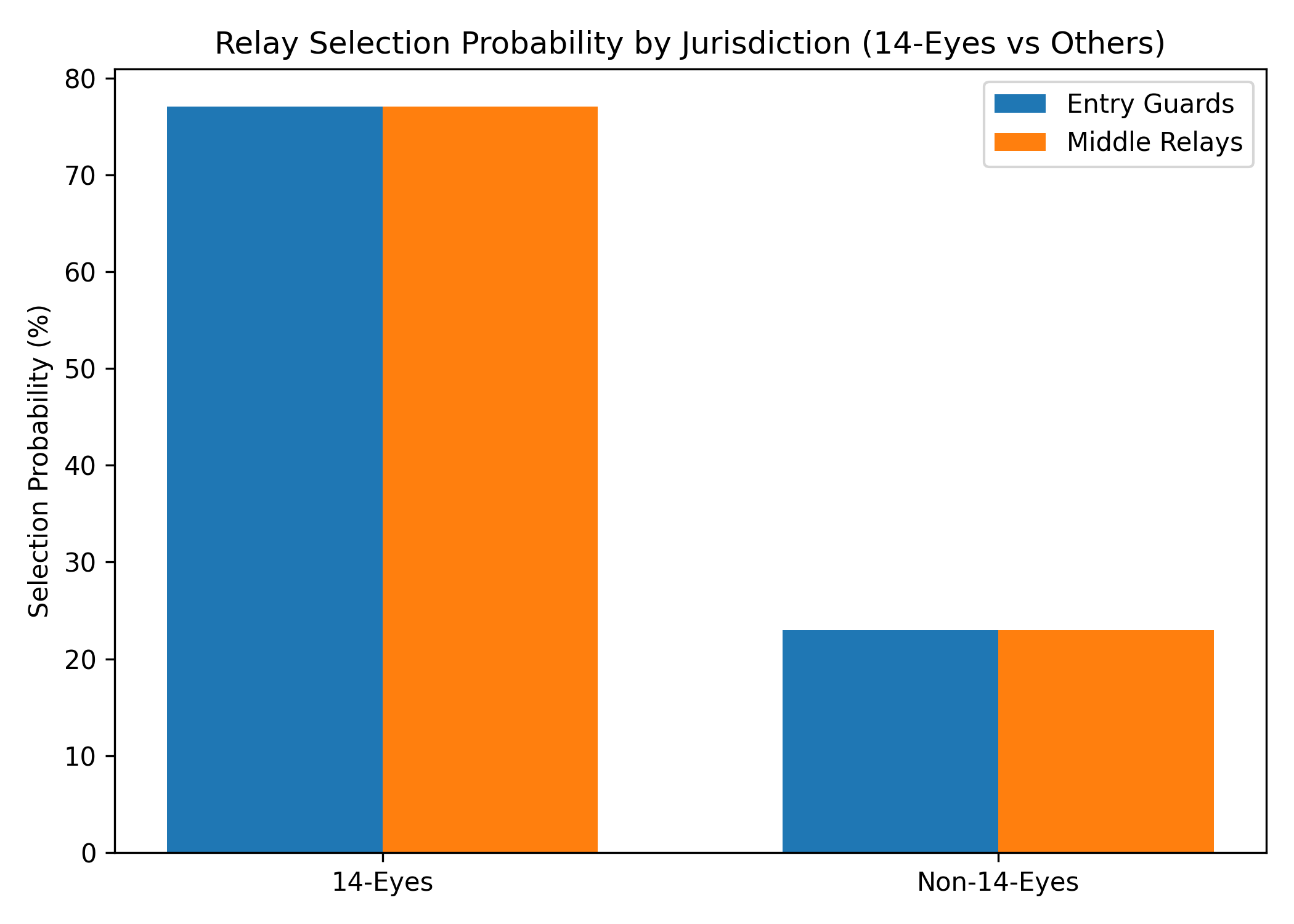}
    \caption{Aggregate guard and middle selection-probability mass inside and outside Fourteen-Eyes jurisdictions, computed from the \texttt{guard\_probability} and \texttt{middle\_probability} values in the 30 November 2025 Onionoo snapshot. Approximately $75\%$ of the probability mass for both positions lies within Fourteen-Eyes jurisdictions.}
    \label{fig:country_probabilities}
\end{figure}

\paragraph{Selection-probability concentration.}
Approximately $75\%$ of both the guard and the middle selection-probability mass
in our snapshot lies within Fourteen-Eyes jurisdictions
(Figure~\ref{fig:country_probabilities}). We obtain this by aggregating the
Onionoo \texttt{guard\_probability} and \texttt{middle\_probability} values of
relays hosted inside and outside those jurisdictions; relay counts are a coarser
measure, because path selection is bandwidth-weighted. As a first indication,
most of the weight at each of these two positions sits inside a single grouping
of jurisdictions whose members already exchange signals.  

\paragraph{Where rotation helps, and where it does not.}
Visibility need not be obtained on the first path the adversary sees. An
introduction circuit is reused for at most $18$--$24$~h
(Section~\ref{sec:onion-service}), and its replacement is built through freshly
chosen relays, so a relay at which visibility could not be obtained may give way
to one at which it can. How far that helps depends on the position; we name each
by its label in Figure~\ref{fig:protocol-percircuit}, where the introduction
circuit runs EG-S $\rightarrow$ VG-S-2 $\rightarrow$ M-S $\rightarrow$ IntP. The
entry guard (EG-S) is the bottleneck: circuits leave through one reachable guard
(Section~\ref{sec:path-selection}), so a rebuild returns to it, and the
adversary must wait for it to fail or for the guard set, held for months, to
shift. Behind it the cost falls away. A rebuild draws a fresh middle relay (M-S)
from the network at large, and redraws the layer-2 vanguard (VG-S-2) from a set
of four that itself turns over every $1$--$12$~days whether or not the adversary
acts (Section~\ref{sec:path-selection}); the Introduction Point (IntP) comes
from the same wide population and changes when the service rotates it. A service's three introduction circuits share that guard, so they multiply the relays
the adversary can try at the other positions and none at that one. The three can
also be attacked concurrently, each reconstruction ending at the same guard and
the same service. Path selection applies no subnet or family rule
here (Section~\ref{sec:predecessor-exclusion}), so nothing resists such
concentration. 
 \FloatBarrier
 %\input{sections/appendix_path_combinations}
 %\FloatBarrier

\end{document}